\documentclass[11pt, letterpaper, onecolumn, oneside, final]{article}

\usepackage[left=1in,right=1in,top=1in,bottom=1in]{geometry}

\usepackage{hyperref}

\usepackage{xcolor}
\usepackage{graphicx}

\usepackage{threeparttable}

\usepackage{amssymb}
\usepackage{amsmath}

\usepackage{tikz}

\usepackage[noline,linesnumbered,boxed,ruled]{algorithm2e}
\usepackage{algpseudocode}

\SetKwBlock{while}{while}{}
\SetKwBlock{s}{At each server $s_i$}{}
\SetKwBlock{w}{At each writer $w_i$}{}
\SetKwBlock{r}{At each reader $r_i$}{}

\SetKwBlock{proc}{procedure}{}
\SetKwBlock{f}{function}{}
\SetKwBlock{upon}{upon}{}
\SetKwBlock{pforeachuntil}{pfor each}{until}
\SetKwBlock{pforeachwaitfor}{pfor each}{wait\,for}
\SetKwFunction{query}{query}
\SetKwFunction{update}{update}
\SetKwFunction{send}{send}
\SetKwFunction{receive}{receive}
\SetKwFunction{TwoRoundWRITE}{TwoRoundWRITE}
\SetKwFunction{OneRoundWRITE}{OneRoundWRITE}
\SetKwFunction{TwoRoundREAD}{TwoRoundREAD}
\SetKwFunction{OneRoundREAD}{OneRoundREAD}
\SetKwFunction{maxSeq}{maxSeq}
\SetKwFunction{maxVer}{maxVer}
\SetKwFunction{valWithMaxVer}{valWithMaxVer}

\SetKwFunction{writers}{write}
\SetKwFunction{readers}{read}

\SetKwFunction{localcache}{localcache}
\SetKwFunction{conflictResolution}{conflictResolution}
\SetKwFunction{DealWithQueryMsg}{DealWithQueryMsg}

\usepackage{amsthm}
\theoremstyle{plain}

\newtheorem{theorem}{Theorem}

\theoremstyle{definition}

\newtheorem{definition}{Definition}[section]

\newcommand{\set}[1]{\left\{#1\right\}}

\newcommand{\pfend}{\hfill $\blacksquare$}

\begin{document}

\title{\bf Fine-grained Analysis on Fast Implementations of Distributed Multi-writer Atomic Registers
}

\author{
  Kaile Huang, Yu Huang, Hengfeng Wei \\
  {\small State Key Laboratory for Novel Software Technology} \\
  {\small Nanjing University} \\
  {\footnotesize mg1933024@smail.nju.edu.cn, \textsf{\{yuhuang, hfwei\}@nju.edu.cn}}
}

\date{}


\maketitle

\begin{abstract}

Distributed multi-writer atomic registers are at the heart of a large number of distributed algorithms. While enjoying the benefits of atomicity, researchers further explore fast implementations of atomic reigsters which are optimal in terms of data access latency. Though it is proved that multi-writer atomic register implementations are impossible when both read and write are required to be fast, it is still open whether implementations are impossible when only write or read is required to be fast. This work proves the impossibility of fast write implementations based on a series of chain arguments among indistiguishable executions. We also show the necessary and sufficient condition for fast read implementations by extending the results in the single-writer case. This work concludes a series of studies on fast implementations of distributed atomic registers.
    
\end{abstract}



\section{Introduction} \label{Sec: Intro}

Distributed storage systems employ replication to improve performance by routing data queries to data replicas nearby \cite{Lakshman10, Redis20, Riak20}. System reliability is also improved due to the redundancy of data. However, data replication is constrained by the intrinsic problem of maintaining data consistency among different replicas \cite{Viotti16}. The data consistency model acts as the ``contract" between the developer and the storage system. Only with this contract can the developers reason about and program over the data items which actually exist as multiple replicas \cite{Sivaramakrishnan15, Burckhardt14}. 

Atomicity is a strong consistency model \cite{Lamport86a, Lamport86b, Herlihy90}. It allows concurrent processes to access multiple replicas of logically the same data item, as if they were accessing one data item in a sequential manner. This abstraction, usually named an \textit{atomic register}, is fundamental in distributed computing and is at the heart of a large number of distributed algorithms \cite{Attiya95, Lynch97}. Though the atomicity model greatly simplifies the development of upper-layer programs, it induces longer data access latency. The latency of read and write operations is mainly decided by the number of round-trips of communications between the reading and writing clients and the server replicas. In the single-writer case, the read operation on an atomic register needs two round-trips of communications \cite{Attiya95}. In the multi-writer case, both write and read operations need two round-trips of communications \cite{Lynch97, Aspnes19}.

In distributed systems, user-perceived latency is widely regarded as the most critical factor for a large class of applications \cite{Lloyd13, Moniz17, Lakshman10, Redis20, Riak20}. While enjoying the benefits of atomicity, researchers further explore whether we can develop \textit{fast} implementations for atomic registers. Since two round-trips are sufficient to achieve atomicity, fast implementation means one round-trip of communication, which is obviously optimal. In the single-writer case, it is proved that when the number of reading clients exceeds certain bound, fast read is impossible \cite{Dutta10}. In the multi-writer case, it is proved impossible when both read and write are required to be fast \cite{Dutta10}. 

This leaves an important open problem when examining the design space of fast implementations of multi-writer atomic registers in a fine-grained manner. Specifically, we denote fast write implementations as W1R2, meaning that the write operation finishes in one round-trip, while the read operation finishes in two round-trips. Similarly, we denote fast read implementations as W2R1 and fast read-write implementations as W1R1. Existing work only proves that fast read-write (W1R1) implementations are impossible. It is still open whether fast write (W1R2) and fast read (W2R1) are impossible. This impossibility result (yet to be proved) underlies the common practice of quorum-replicated storage system design, e.g. the Cassandra data store \cite{Lakshman10}: when read or write is required to finish in one round-trip, weak consistency has to be accepted.

This work thoroughly explores the design space of fast implementations of multi-writer atomic registers. Specifically, for fast write (W1R2) implementations, we prove that it is impossible to achieve atomicity. The impossibility proof is mainly based on the chain argument to construct the indistinguishability between executions. Unlike the W1R1 case, the chain argument for W1R2 implementations faces two severe challenges:
\begin{itemize}
    \item (Section 3) Since the read operation has one more round-trip (compared to the W1R1 case) to discover differences between executions, it is more difficult to construct the indistinguishability we need for the impossibility proof. To this end, we combine three consecutive rounds of chain arguments, in order to hide the differences in the executions from the 2-round-trip read operations.
    
    \item (Section 4) The first round-trip of a read operation might update information on the servers, thus potentially affecting the return values of other read operations. The effect from the first round-trip of read operations may also break the indistinguishability we try to construct. To this end, we use sieve-based construction of executions to eliminate the effect of the first round-trip of a read operation.
    
\end{itemize}
        
\noindent The impossibility proof for W1R2 implementations is the main contribution of this work. 

For W2R1 implementations, we prove the impossibility when $R \geq \frac{S}{t} - 2$. When $R < \frac{S}{t} - 2$, we propose a W2R1 implementation. The proof and the implementation are extensions to the results of the single-writer case \cite{Dutta10}. 

The contributions in this work conclude a series of studies on fast implementations of distributed atomic registers. The contributions of this work in light of results in the existing work are outlined in Table \ref{T: Contri}.

\begin{table}[t]
\caption{Overview of contributions.}
\vspace{0.1cm}
\label{T: Contri} 
\centering
\begin{tabular}{l | l | l}
    \hline
    Design space & Impossibility & Implementation \\
    \hline
    \hline

   
   W2R2 $^{\scriptsize{\cite{Lynch97}}}$ & $t \geq \frac{S}{2}$ & $W \geq 2, R\geq 2, t < \frac{S}{2}$ \\ \hline
    

    W1R2 $^{\scriptsize{[\text{this work}]}}$ & $W \geq 2, R\geq 2, t\geq 1$ & $\emptyset$ \\ \hline

    W2R1 $^{\scriptsize{[\text{this work}]}}$ & $R \geq \frac{S}{t} - 2$ & $R < \frac{S}{t} - 2$ \\ \hline
    
    W1R1 $^{\scriptsize{\cite{Dutta10}}}$ & $W \geq 2, R\geq 2, t\geq 1$ & $\emptyset$ \\
    
    \hline
\end{tabular}
\end{table}


The rest of this work is organized as follows. In Section \ref{Sec: Preli}, we describe the preliminaries. Section \ref{Sec: W1R2-Chain} and Section \ref{Sec: W1R2-Sieve} present the impossibility proof for W1R2 implementations. Section \ref{Sec: W2R1} outlines the impossibility proof and the algorithm design of W2R1 implementations. Section \ref{Sec: RW} discusses the related work. In Section \ref{Sec: Concl}, we conclude this work and discuss the future work.

\section{Preliminaries} \label{Sec: Preli}

In this section, we first describe the system model and the definition of atomicity. Then we outline the algorithm schema for multi-writer atomic register implementations.

\subsection{Atomic Register Emulation in Message-passing Systems} \label{Subsec: Model}

We basically adopt the system model used in \cite{Dutta10}. Specifically, a replicated storage system considered in this work consists of three disjoint sets of processes: 

 
\begin{itemize}
    \item the set $\Sigma_{sv}$ of \textit{servers}: $\Sigma_{sv} = \set{s_1, s_2, \cdots, s_S}$.
    \item the set $\Sigma_{rd}$ of \textit{readers}: $\Sigma_{rd} = \set{r_1, r_2, \cdots, r_R}$.
    \item the set $\Sigma_{wr}$ of \textit{writers}: $\Sigma_{wr} = \set{w_1, w_2, \cdots, w_W}$.  
\end{itemize}

\begin{figure}[ht]
    \centering
    \includegraphics[width = 0.9\columnwidth ]{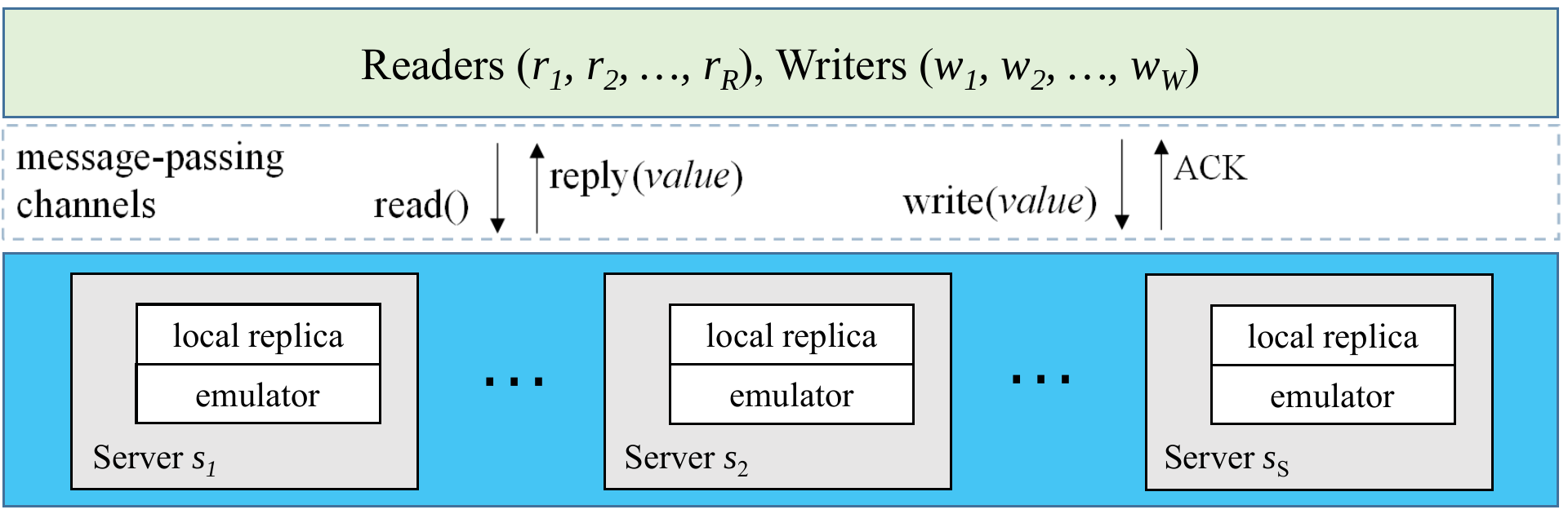}
    \caption{System model of read/write register emulation.}
    \label{F: System-Overview}
\end{figure}

\noindent Here, $S$, $R$ and $W$ denote the cardinalities of $\Sigma_{sv}$, $\Sigma_{rd}$ and $\Sigma_{wr}$ respectively. The readers and the writers are also called \textit{clients}. We are concerned of \textit{multi-writer multi-reader} implementations. Thus we have $W\geq 2$ and $R\geq 2$. In a distributed message-passing system, we also have that $S \geq 2$. The clients and the servers communicate by asynchronous message passing, via a bidirectional reliable communication channel, as shown in Fig. \ref{F: System-Overview}. There is no communication among the servers. For the simplicity of presentation, we assume the existence of a discrete global clock, but the processes cannot access the global clock. An implementation $\mathcal{A}$ of a shared register is a collection of automata. Computation proceeds in steps of $\mathcal{A}$. An execution is a finite sequences of steps of $\mathcal{A}$. In any given execution, any number of readers and writers, and $t$ out of $S$ servers may crash.

An atomic register is a distributed data structure that may be concurrently accessed by multiple clients, yet providing an ``illusion of a sequential register" to the accessing processes. The atomic register provides two types of operations. Only a writer can invoke the write operation $write(v)$, which stores $v$ in the register. Only a reader can invoke the read operation $read()$, which returns the value stored. We are concerned of \textit{wait-free} implementations, where any read or write invocation eventually returns independently of the status of other clients. Due to the locality property of atomicity \cite{Herlihy90}, we consider one single shared register.

We define an \textit{execution} of the clients accessing the shared register as a sequence of events where each event is either the invocation or the response of a read or write operation.  Each event in the execution is tagged with a unique timestamp from the global clock, and events appear in the execution in increasing order of their timestamps. For execution $\sigma$, we can define the partial order between operations. Let $O.s$ and $O.f$ denote the timestamps of the invocation and the response events of operation $O$ respectively. We define $O_1 \prec_\sigma O_2$ if $O_1.f < O_2.s$. We define $O_1 || O_2$ if neither $O_1 \prec_\sigma O_2$ nor $O_2 \prec_\sigma O_1$ holds. An execution $\sigma$ is \textit{sequential} if $\sigma$ begins with an invocation, and each invocation is immediately followed by its matching response. An execution $\sigma$ is \textit{well-formed} if for each client $p_i$, $\sigma|p_i$ (the subsequence of $\sigma$ restricted on $p_i$) is sequential. Given the notations above, we can define atomicity:
\begin{definition} \label{Def: Atomicity}

A shared register provides \textit{atomicity} if, for each of its well-formed executions $\sigma$, there exists a permutation $\pi$ of all operations in $\sigma$ such that $\pi$ is sequential and satisfies the following two requirements:
    \begin{itemize}
        \item {$[$Real-time requirement$]$ } If $O_1 \prec_\sigma O_2 $, then $O_1$ appears before $O_2$ in $\pi$.
        
        \item {$[$Read-from requirement$]$ } Each read returns the value written by the latest preceding write in $\pi$.
    \end{itemize}
    
\end{definition}

\subsection{Algorithm Schema for Multi-writer Atomic Register Implementations} \label{Subsec: Diamond}

When studying fast implementations of multi-writer atomic registers, the critical operation we consider is the round-trip of communication between the client and the servers. In each round-trip, the client can \textit{query} all the servers, i.e., collect useful information from the servers. The client can also \textit{update} all the servers, i.e., send useful information to the servers. Upon receiving a \textit{query} request, the server replies the client as required. Upon receiving an \textit{update} request, the server first stores data sent from the client. Then it can reply certain information if necessary, or it can simply reply an ACK. Exemplar implementations can be found in \cite{Attiya95, Lynch97, Aspnes19, Wei17, Huang20}.

\begin{figure}[ht]
    \centering
    \includegraphics[width = 0.5\columnwidth ]{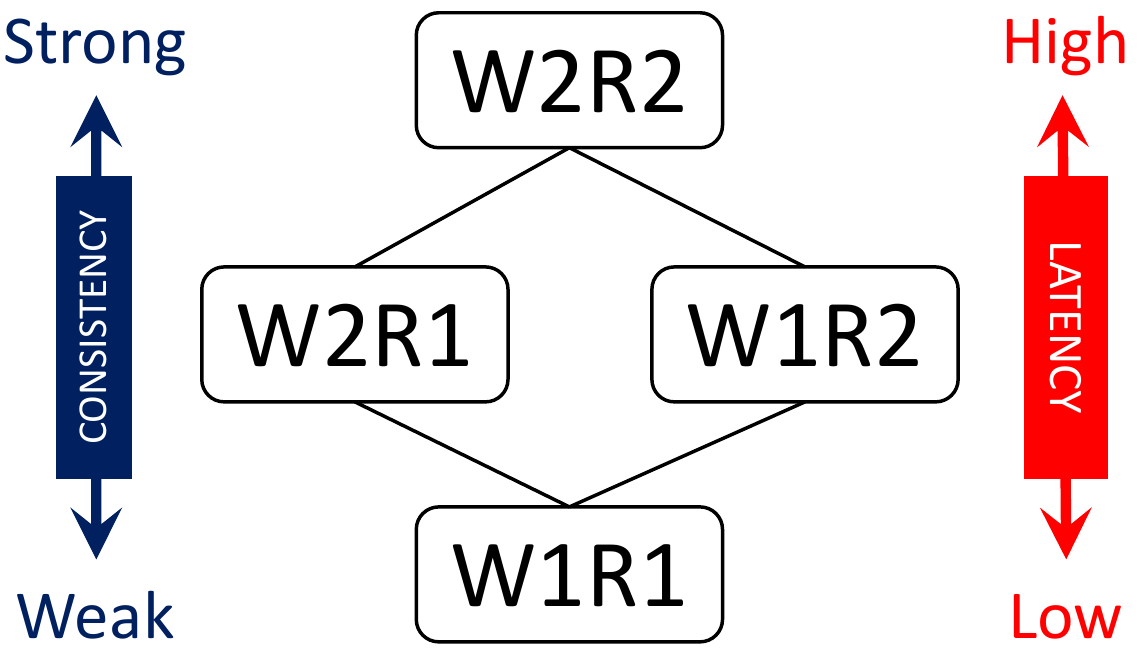}
    \caption{Algorithm schema for multi-writer atomic register implementations.}
    \label{F: Diamond}
\end{figure}

Tuning the number of round-trips in emulation of a multi-writer atomic register, we have four possible types of implementations \cite{Ouyang19}, as shown in Fig. \ref{F: Diamond}. They are slow read-write implementation (W2R2), fast write implementation (W1R2), fast read implementation (W2R1) and fast read-write implementation (W1R1). Fig. \ref{F: Diamond} can be viewed as the Hasse Diagram of the partial order among implementations. The partial order relation can be thought of as providing stronger consistency guarantees or inducing less data access latency.

Note that, for atomic register implementations, when 2 round-trips are sufficient, we do not consider implementations employing $k$ round-trips for $k \geq 3$. However, for impossibility of fast implementations, we need to consider the impossibility of W1R$k$ and W$k$R1 implementations for $k\geq 3$. The impossibility proofs of W1R$k$ and W$k$R1 implementations are principally same with the impossibility proofs of W1R2 and W2R1 implementations, as discussed in Section \ref{Sec: W1R2-Chain} and Section \ref{Sec: W2R1} respectively.

\section{Fast Write (W1R2): Chain Arguments for Impossibility Proof} \label{Sec: W1R2-Chain}

We first present the impossibility proof for fast write (W1R2) implementations. Specifically, we prove the following theorem: 

\begin{theorem} [W1R2 impossibility] \label{Thm: W1R2-Impo}
    Let $t \geq 1$, $W \geq 2$ and $R \geq 2$. There is no fast write \emph{(W1R2)} atomic register implementation.
\end{theorem}

\noindent This impossibility result is proved by chain argument \cite{Attiya14}, which is also used to prove the impossibility of W1R1 implementations in \cite{Dutta10}. The central issue in chain argument is to construct certain indistinguishability between executions. Compared to the impossibility proof of W1R1 implementations, the read operations now have one more round-trip. This ``one more round-trip" imposes two critical challenges for constructing the indistinguishability:
\begin{enumerate}
    \item Obtaining more information from the second round-trip, the read operations can now ``beat" the indistinguishability constructed in the W1R1 case. In our proof, we add one more read operation and construct two more chains of executions, in order to obtain the indistinguishability even when facing two round-trips of read operations.
    
    \item The first round-trip of a read operation may update information on the servers, thus possibly affecting the return values of other read operations. The effect of the first round-trip may also break the indistinguishability we plan to construct. To cope with this challenge, we propose the sieve-based construction of executions. We sieve all the servers and eliminate those which are affected by the first round-trip of a read operation. On the servers that remain after the sieving, we show that the chain argument can still be successfully conducted.
\end{enumerate}

\noindent This section addresses the first challenge and presents the chain argument. In Section \ref{Sec: W1R2-Sieve}, we address the second challenge and discuss how to eliminate effects of the first round-trip. Note that the impossibility proof of W1R2 implementations also applies for W1R$k$ implementations for $k \geq 3$. We can combine the round-trips $2, 3, \cdots, k$ as if they were one single round-trip. The chain argument still applies.

\subsection{Overview} 

It suffices to show the impossibility in a system where $S\geq 3$ \footnote[2]{In a replicated system, we have $S\geq 2$. When $S=2$ while $t=1$, it is trivial to prove the impossibility.}, $W = 2$, $R = 2$ and $t = 1$. In the proof, we use two write operations $W_1$ and $W_2$ (issued by writers $w_1$ and $w_2$ respectively) and two read operations $R_1$ and $R_2$ (issued by readers $r_1$ and $r_2$ respectively). Since $t=1$, the read operation must be able to return when one server gives no response. When constructing an execution, we say one round-trip in an operation \textit{skips} one server $s$, if the messages between the client and the server are delayed a sufficiently long period of time (e.g. until the rest of the execution has finished). If one round-trip of communication does not skip any server, we say it is \textit{skip-free}.

In a chain argument, we will construct a chain of executions, where two consecutive executions in the chain differ only on one server. Since in two end executions of the chain the read operations return different values, there must be some ``critical server". The change on the critical server results in the difference in the return values. We intentionally let the read operation skip the critical server. This will construct the indistinguishability we need (as detailed in Section \ref{SubSec: Phase1}). 

In a chain argument, we may also utilize the relation between operations to construct the indistinguishability (as detailed in Section \ref{SubSec: Phase3}). Specifically, one operation cannot notice the differences in executions after it has finished. Moreover, the operation cannot notice the differences  on one server if it skips this server. 

The indistinguishability makes the read operations return the same value in two executions. However, construction of the chain of executions tells us that the two executions should return different values (note that within one execution, two reads must return the same value, as required by the definition of atomicity). This leads to contradiction. 

\begin{figure*}[htb]
	\center
	\includegraphics[width=0.95\linewidth]{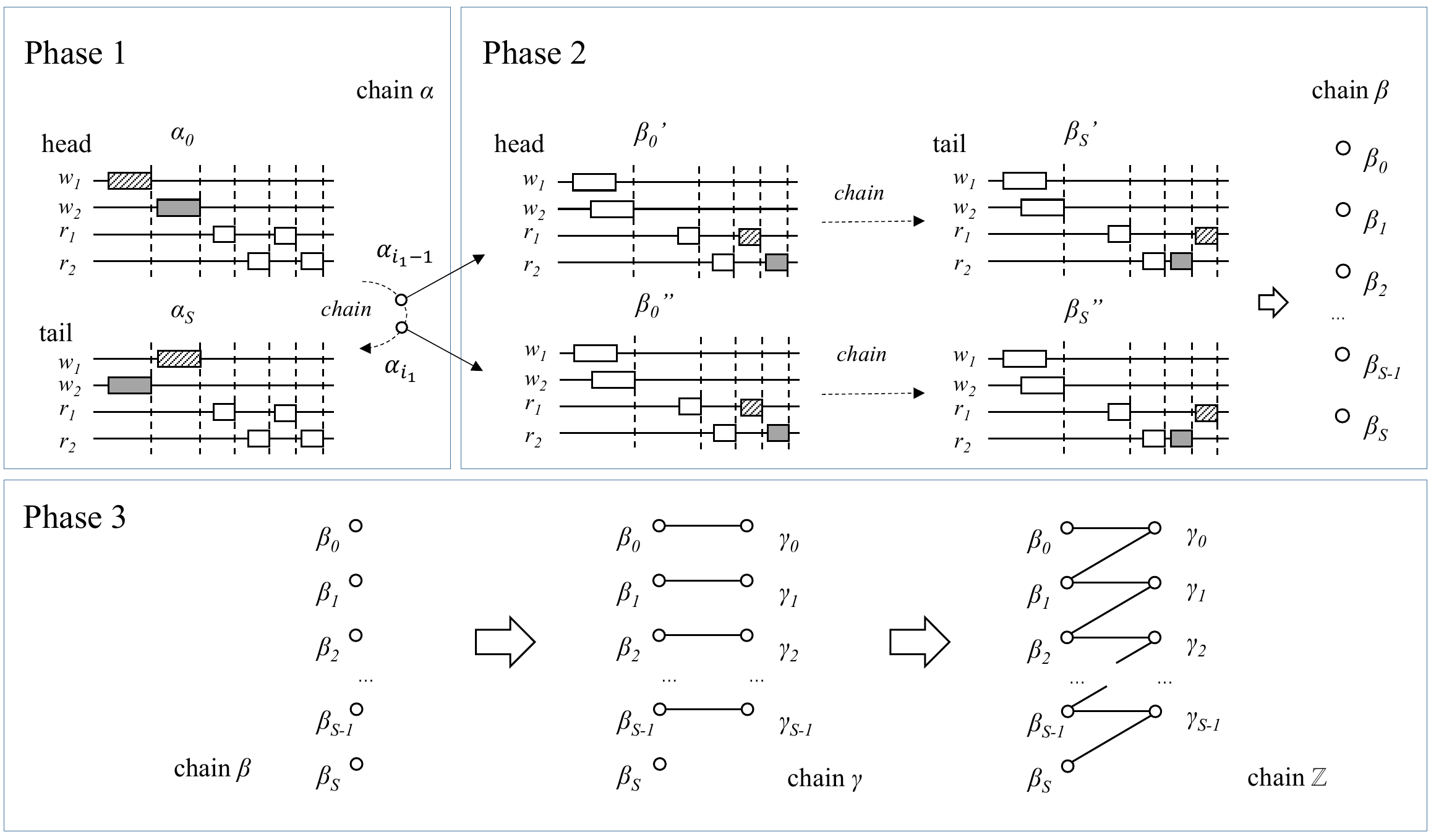}
	\caption{Proof overview.}
	\label{F: Proof-Overview}
\end{figure*}

To beat the ability of the read operation to employ two round-trips of communications, we need to conduct a series of chain arguments. The proof will be presented in three phases, each phase constructing one chain, as shown in Fig. \ref{F: Proof-Overview}. For the ease of presentation, we assume in the chain argument that the first round-trip of a read operation will not affect the return values of other read operations. In Section \ref{Sec: W1R2-Sieve}, we will explain how to lift this assumption.

\subsection{Phase 1: Chain $\alpha$ and Critical Server $s_{i_1}$} \label{SubSec: Phase1}

To construct chain $\alpha$, we first construct the ``head" and the ``tail" executions:
\begin{itemize}
    \item Head execution $\alpha_{head}$ consists of the following three non-concurrent operations: i) a skip-free $W_1 = write(1)$, which precedes ii) a skip-free $W_2 = write(2)$, which precedes iii) a skip-free $R_1 = read()$. Note that all servers receive the three operations in this order. In $\alpha_{head}$, $R_1$ returns 2 (as required by the definition of atomicity).
    
    \item Tail execution $\alpha_{tail}$ consists of the same three non-concurrent operations, but the temporal order is $W_2$, $W_1$ and $R_1$. In $\alpha_{tail}$, $R_1$ returns 1.
\end{itemize} 

\noindent Let $\alpha_0 = \alpha_{head}$. From $\alpha_0$ we construct $\alpha_1$, the next execution in the chain, as follows. Execution $\alpha_1$ is identical to $\alpha_0$ except that server $s_1$ receives $W_2$ first, and then $W_1$ and $R_1$. That is, we ``swap" two write operations on $s_1$ and everything else is unchanged. Continuing this ``swapping" process, we swap two write operations on $s_i$ in $\alpha_{i-1}$ and obtain $\alpha_i$, for all $1\leq i \leq S$. Thus we obtain chain $\alpha = (\alpha_0, \alpha_1, \cdots, \alpha_S)$. Note that $R_1$ cannot distinguish $\alpha_S$ from $\alpha_{tail}$. Thus, $R_1$ returns 1 in $\alpha_S$, while it returns 2 in $\alpha_0$.

Since $R_1$ returns different values in two ends of the chain, there must exist two consecutive executions $\alpha_{i_1-1}$ and $\alpha_{i_1}$ ($1\leq i_1 \leq S$), such that $R_1$ returns 2 in $\alpha_{i_1-1}$ and returns 1 in $\alpha_{i_1}$. Note that $\alpha_{i_1-1}$ and $\alpha_{i_1}$ differ only on one ``critical server" $s_{i_1}$, i.e., $s_{i_1}$ receives $W_1$ first in $\alpha_{i_1-1}$,  and receives $W_2$ first in $\alpha_{i_1}$. This critical server $s_{i_1}$ will be intentionally skipped to obtain indistinguishability, when constructing chain $\beta$ in Phase 2 below. 

\subsection{Phase 2: Chain $\beta$ Derived from Chain $\beta'$ and Chain $\beta''$} \label{SubSec: Phase2}

In Phase 2 of our proof, we basically append the second read operation $R_2$ to executions in chain $\alpha$ and obtain chain $\beta$. We actually construct two candidate chains $\beta'$ and $\beta''$, and modify one of them to get chain $\beta$, depending on what the return value of $R_2$ is. Chain $\beta'$ and $\beta''$ stem from execution $\alpha_{i_1-1}$ and $\alpha_{i_1}$ respectively, i.e. two executions pertained to the critical change on the critical server.

Since the read operations consist of two round-trips, we denote the two round-trips of read operation $R_i$ as $R_i^{(1)}$ and $R_i^{(2)}$ ($i=1, 2$). We extend execution $\alpha_{i_1-1}$ with the second read operation $R_2$. We interleave the round-trips of $R_1$ and $R_2$ as follows: the four round-trips are non-concurrent and the temporal order is $R_1^{(1)}$, $R_2^{(1)}$, $R_1^{(2)}$ and $R_2^{(2)}$ on all servers $s_i$ ($1\leq i \leq S$), as shown in Fig. \ref{F: Proof-Overview}. This execution is named $\beta'_{head} = \beta'_0$. To construct chain $\beta'$, we will swap $R_1^{(2)}$ and $R_2^{(2)}$ on one server a time. Specifically, for $1\leq i \leq S$, $\beta'_i$ is the same with $\beta'_{i-1}$, except that server $s_i$ receives $R_1^{(2)}$ first in $\beta'_{i-1}$, and receives $R_2^{(2)}$ first in $\beta'_i$. The last execution of the chain is $\beta'_{tail} = \beta'_S$.

We then extend execution $\alpha_{i_1}$ in the same way, and get $\beta''_{head} = \beta''_0$. We also do the swapping in the same way and get executions $\beta''_1, \beta''_2, \cdots, \beta''_S$. The only difference between chain $\beta'$ and $\beta''$ is that, chain $\beta'$ stems from  execution $\alpha_{i_1-1}$, while chain $\beta''$ stems from $\alpha_{i_1}$. Thus, $R_1$ returns 2 in chain $\beta'$, while returning 1 in chain $\beta''$. This is because, the return value of $R_1$ is decided by executions $\alpha_{i_1-1}$ and $\alpha_{i_1}$. Appending the read operation $R_2$ should not change the return value of an existing read, as required by the definition of atomicity.

The only server which can tell the difference between $\beta'$ and $\beta''$ is $s_{i_1}$, the critical server in chain $\alpha$. Now we modify tail executions $\beta'_{tail}$ and $\beta''_{tail}$, in order to obtain the indistinguishability we need. In both tail executions $\beta'_{tail}$ and $\beta''_{tail}$, we let $R_2$ (both round-trips) skip server $s_{i_1}$. Thus the (modified) $\beta'_{tail}$ and $\beta''_{tail}$ are indistinguishable to $R_2$, and $R_2$ returns the same value in both modified tail executions. 

To construct chain $\beta$, we must start from either $\beta'$ or $\beta''$, and revise the chosen candidate chain into chain $\beta$. The criteria for choosing a chain is that the candidate chain must enable us to make the read operations in the two 	end executions $\beta_0$ and $\beta_S$ have different return values. Without loss of generality, we assume that $R_2$ returns 1 in both $\beta'_{tail}$ and $\beta''_{tail}$ (modified, with $R_2$ skipping $s_{i_1}$). In $\beta'_0$, since $R_1$ returns 2, according to the definition of atomicity, we have that $R_2$ must also return 2 in $\beta'_0$. Thus, we choose chain $\beta'$. 

We modify chain $\beta'$ to obtain chain $\beta$ as follows. For every execution in chain $\beta'$, we let $R_2$ (both round-trips) skip $s_{i_1}$ and obtain every corresponding execution in chain $\beta$. That is, $R_2$ in chain $\beta'$ is skip-free while $R_2$ in chain $\beta$ skips $s_{i_1}$ (if $R_2$ returns 2 in both modified $\beta'_{tail}$ and $\beta''_{tail}$, we will choose to revise chain $\beta''$, and obtain chain $\beta$ in the same way). Chain $\beta$ servers as the basis for construction of chain $\gamma$ and $\mathbb{Z}$ in Phase 3 of our proof.

\subsection{Phase 3: Zigzag Chain $\mathbb{Z}$ Combining Chain $\beta$ and $\gamma$} \label{SubSec: Phase3}

Given chain $\beta$, we have that $R_1$ and $R_2$ both return 2 in $\beta_0$, while both read operations return 1 in $\beta_S$. Now in Phase 3, we will first construct chain $\gamma = (\gamma_0, \gamma_1, \cdots, \gamma_{S-1})$. Then we combine chain $\beta$ and $\gamma$, and obtain the zigzag chain $\mathbb{Z}$, as shown in Fig. \ref{F: Proof-Overview}.

For any two executions $x$	and $x'$ from chain $\beta$ and $\gamma$, we define an equivalence relation: $x \approx x'$ when $R_1$ and $R_2$ return the same value in both $x$ and $x'$. Note that $R_1$ and $R_2$ must return the same value in one execution, as required by the definition of atomicity. We will prove that all executions in chain $\mathbb{Z}$ are connected by the `$\approx$' relation, i.e., $\beta_0 \approx \gamma_0 \approx \beta_1 \approx \gamma_1 \approx \cdots \beta_{S-1} \approx \gamma_{S-1} \approx \beta_S$. According to our construction in Phase 1 and 2, we have that $\beta_0 \not\approx \beta_S$. This leads to contradiction.

We first construct the horizontal links in chain $\mathbb{Z}$, i.e., $\forall 0\leq k \leq S-1, \beta_k \approx \gamma_k$ in Section \ref{SubsubSec: H-Link}. Then we construct the diagonal links, i.e., $\forall 0\leq k \leq S-1, \beta_{k+1} \approx \gamma_k$ in Section \ref{SubsubSec: D-Link}.

\subsubsection{Horizontal link from $\beta_k$ to $\gamma_k$} \label{SubsubSec: H-Link}

We first construct execution $\gamma_k$ from $\beta_k$ ($0\leq k \leq S-1$). The construction process implies that $\beta_k \approx \gamma_k$. The key behind the process is still constructing certain indistinguishability. When constructing $\gamma_i$, we need to utilize two sources of indistinguishability: 
\begin{enumerate}
    \item When $R_1^{(2)}$ finishes before $R_2^{(2)}$ on some server $s_x$, and we modify $R_2^{(2)}$ on $s_x$, $R_1^{(2)}$ will not notice the change (behind its back).
    
    \item When $R_2^{(2)}$ skips $s_x$, and we modify $R_1$ on $s_x$, $R_2$ will not notice the change.
\end{enumerate}

\begin{figure}[ht]
    \center
    \includegraphics[width=0.7\columnwidth]{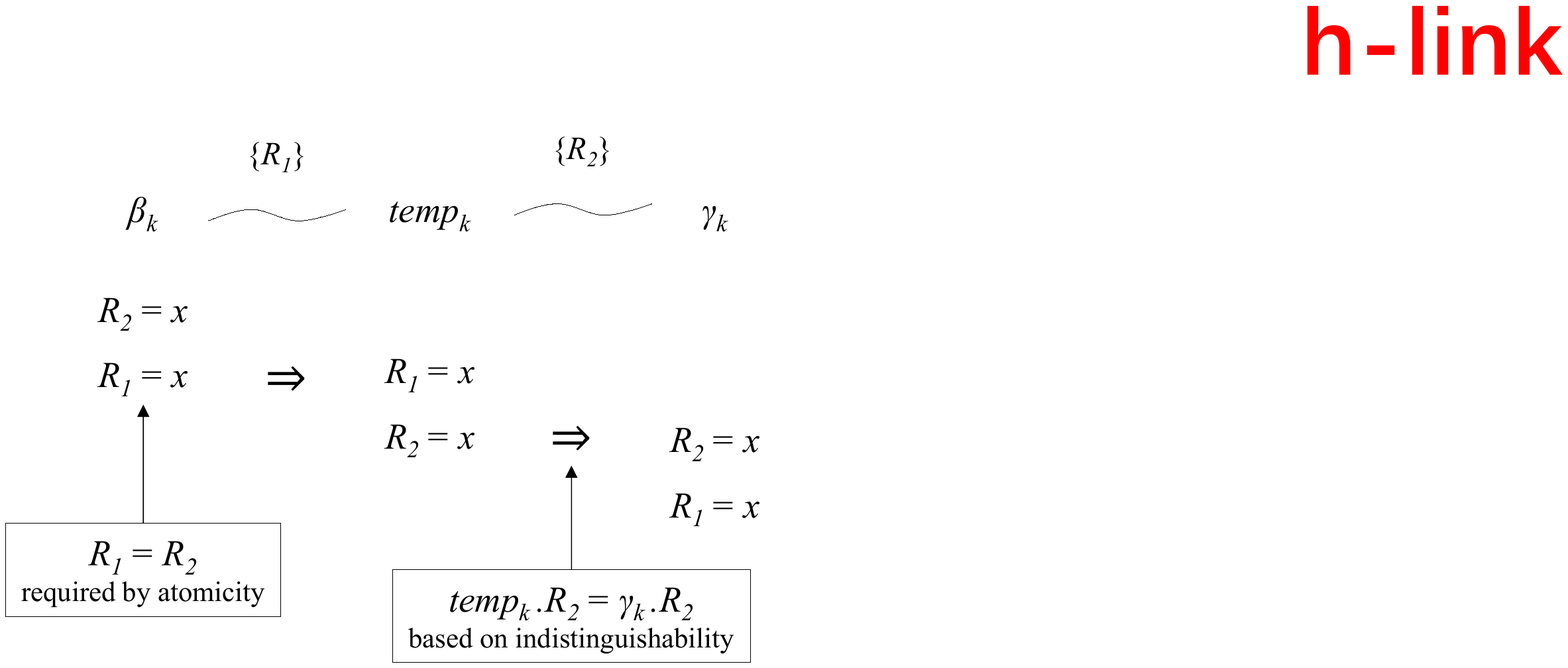}
    \caption{Construction of the horizontal link: the reader's view.}
    \label{F: H-Link1}
\end{figure}

\begin{figure}[ht]
    \center
    \includegraphics[width=\columnwidth]{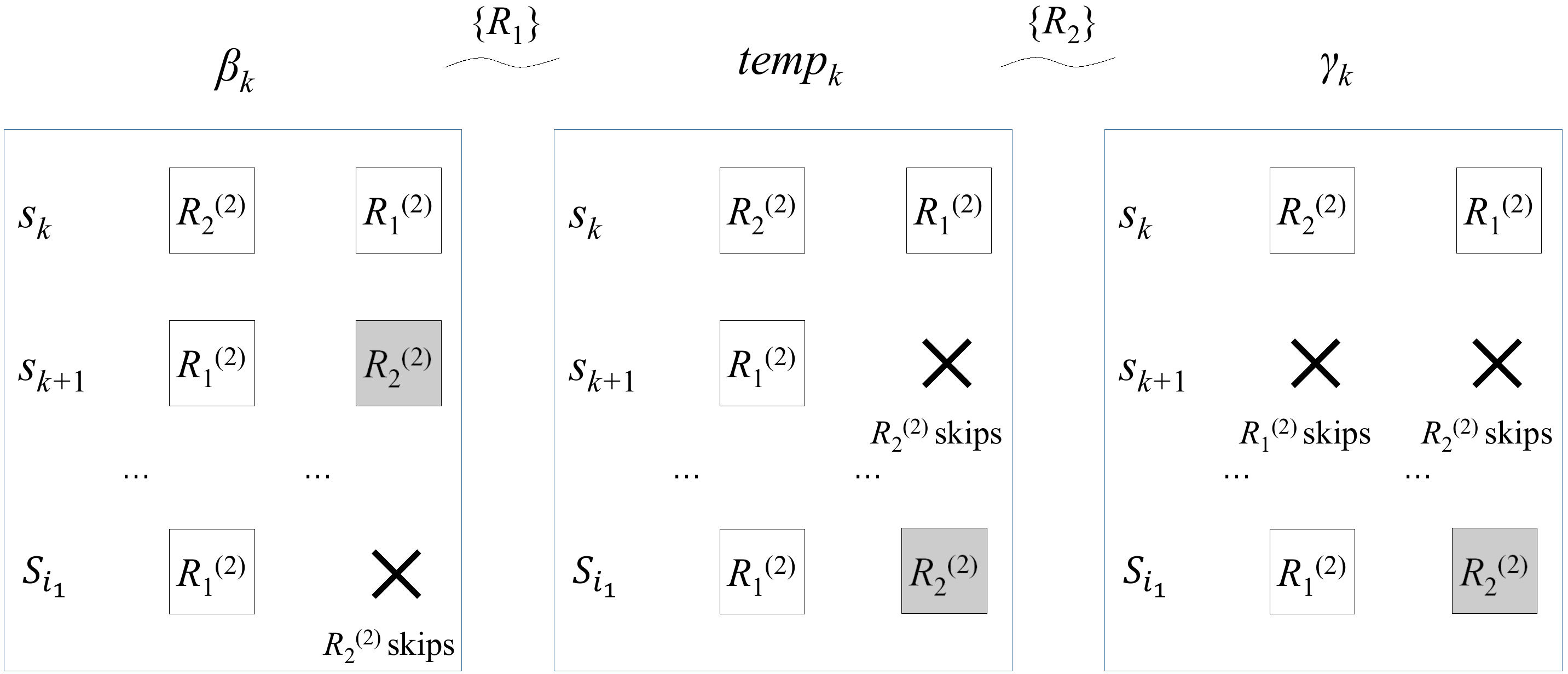}
    \caption{Construction of the horizontal link: the server's view.}
    \label{F: H-Link2}
\end{figure}

\noindent The construction is shown in Fig. \ref{F: H-Link1} and Fig. \ref{F: H-Link2}, from the reader's view and the server's view respectively.

Before the construction of $\gamma_k$, we need to review the characteristics of all executions in chain $\beta$. For every execution $\beta_k$ ($0 \leq k \leq S$), operation $R_1$ (both round-trips) is skip-free, while $R_2$ (both round-trips) skips exactly one server $s_{i_1}$ (the critical server obtained from chain $\alpha$, see Section \ref{SubSec: Phase1}). In the construction of $\gamma_k$, we will change the server $R_2^{(2)}$ skips. We will also let $R_1^{(2)}$ skip one server. No other modifications will be made to $\beta_k$. Also note that starting from $\beta_0$, for $k=1, 2, \cdots, S$, we do the swapping in $s_k$ and obtain execution $\beta_k$. That is, for $\beta_k$, $s_{k+1}$ sees $R_1^{(2)}$ and then $R_2^{(2)}$ (not swapped); while $s_k$ sees $R_2^{(2)}$ and then $R_1^{(2)}$ (swapped).

From $\beta_k$, we will create $\gamma_k$ as follows. In construction of $\gamma_k$, we only modify $R_1^{(2)}$ and  $R_2^{(2)}$, i.e., the first round-trips of both operations are unchanged. In $\beta_k$, the swapping takes place on $s_k$ and we will pick the first not-swapped server, i.e., $s_{k+1}$. Server $s_{k+1}$ finishes $R_1^{(2)}$ before it receives $R_2^{(2)}$. We create a temporary execution $temp_k$ which is the same with $\beta_k$ except that $R_2^{(2)}$ skips $s_{k+1}$ and does not skip $s_{i_1}$. The only two servers affected are $s_{k+1}$ and $s_{i_1}$. Note that here we assume that $k+1\neq i_1$. The case $k+1=i_1$ (which is actually simpler) will be discussed separately below. For the two servers affected, we verify the indistinguishability for $R_1$ (see Fig. \ref{F: H-Link2}):
\begin{itemize}
    \item For $s_{k+1}$, $R_1$ cannot see any difference since $R_1$ finishes first.
    
    \item For $s_{i_1}$, previously $R_2^{(2)}$ skips $s_{i_1}$ (in $\beta_k$) and now we add $R_2^{(2)}$ back on $s_{i_1}$ (in $temp_k$). We can intentionally add $R_2^{(2)}$ after $R_1^{(2)}$ on $s_{i_1}$. Thus $R_1$ still cannot see any difference.
\end{itemize}

\noindent Thus $R_1$ cannot distinguish $\beta_k$ from $temp_k$, and $R_1$ will return the same value in both executions (see Fig. \ref{F: H-Link1}). As required by the definition of atomicity, $R_2$ will return the same value with $R_1$, thus returning the same value in both executions. This gives us that $\beta_k \approx temp_k$.

Now we create execution $\gamma_k$ which is the same with $temp_k$ except that $R_1^{(2)}$ skips $s_{k+1}$ (note that in $\beta_k$ and $temp_k$, $R_1^{(2)}$ is skip-free). The only change takes place on $s_{k+1}$. Since $R_2^{(2)}$ skips $s_{k+1}$ in both $temp_k$ and $\gamma_k$, $R_2$ cannot distinguish $temp_k$ from $\gamma_k$. Thus we have that $R_2$ will return the same value in $temp_k$ and $\gamma_k$ (see Fig. \ref{F: H-Link1}). Also as required by the definition of atomicity, $R_1$ will return the same value in both executions. This gives us $temp_k \approx \gamma_k$.

Finally, combining the two links above (see Fig. \ref{F: H-Link1} and Fig. \ref{F: H-Link2}), we have $\beta_k \approx \gamma_k$. Here note that since $R_2^{(2)}$ skips $s_{k+1}$, it seems unnecessary for $R_1$ to skip $s_{k+1}$ in $\gamma_k$. For the proof till now, it is indeed unnecessary. However, we need to let $R_1$ skip $s_{k+1}$ here, in order to construct the diagonal link between $\beta_{k+1}$ and $\gamma_k$ later in the following Section \ref{SubsubSec: D-Link}.

In the proof above, we left out the case $k+1 = i_1$, which is discussed here. When $k+1=i_1$, we create $\gamma_k$ as follows. In $\beta_k$, $s_{k+1}$ only receives $R_1^{(2)}$ (since $R_2^{(2)}$ skips $s_{i_1} = s_{k+1}$). We let $R_1^{(2)}$ skip $s_{k+1}$, and get $\gamma_k$. Since $R_2^{(2)}$ skips $s_{k+1}$, $R_2$ cannot distinguish $\beta_k$ from $\gamma_k$ and will return the same value. As required by the definition of atomicity, $R_1$ will also return the same value in $\beta_k$ and $\gamma_k$. Thus we still have $\beta_k \approx \gamma_k$ when $k+1=i_1$.

\subsubsection{Diagonal link from $\beta_{k+1}$ to $\gamma_k$} \label{SubsubSec: D-Link}

Now we construct the diagonal link. We will create from $\beta_{k+1}$ executions $temp_k'$ and $\gamma_k'$ ($0\leq k \leq S-1$). The construction is principally the same with the construction of the horizontal link. We need to show that $\beta_{k+1} \approx temp_k' \approx \gamma_k'$. As for $\gamma_k'$ and $\gamma_k$, the executions on all servers, together with the order among operations, are the same. It is straightforward to verify that $\gamma_k' \approx \gamma_k$ (so we do not show $\gamma_k'$ in Phase 3 in Fig. \ref{F: Proof-Overview}). Thus we can obtain the diagonal link, meaning that $\beta_{k+1} \approx \gamma_k$. Now we explain construction of the diagonal link in detail.

\begin{figure}[ht]
    \center
    \includegraphics[width=0.7\columnwidth]{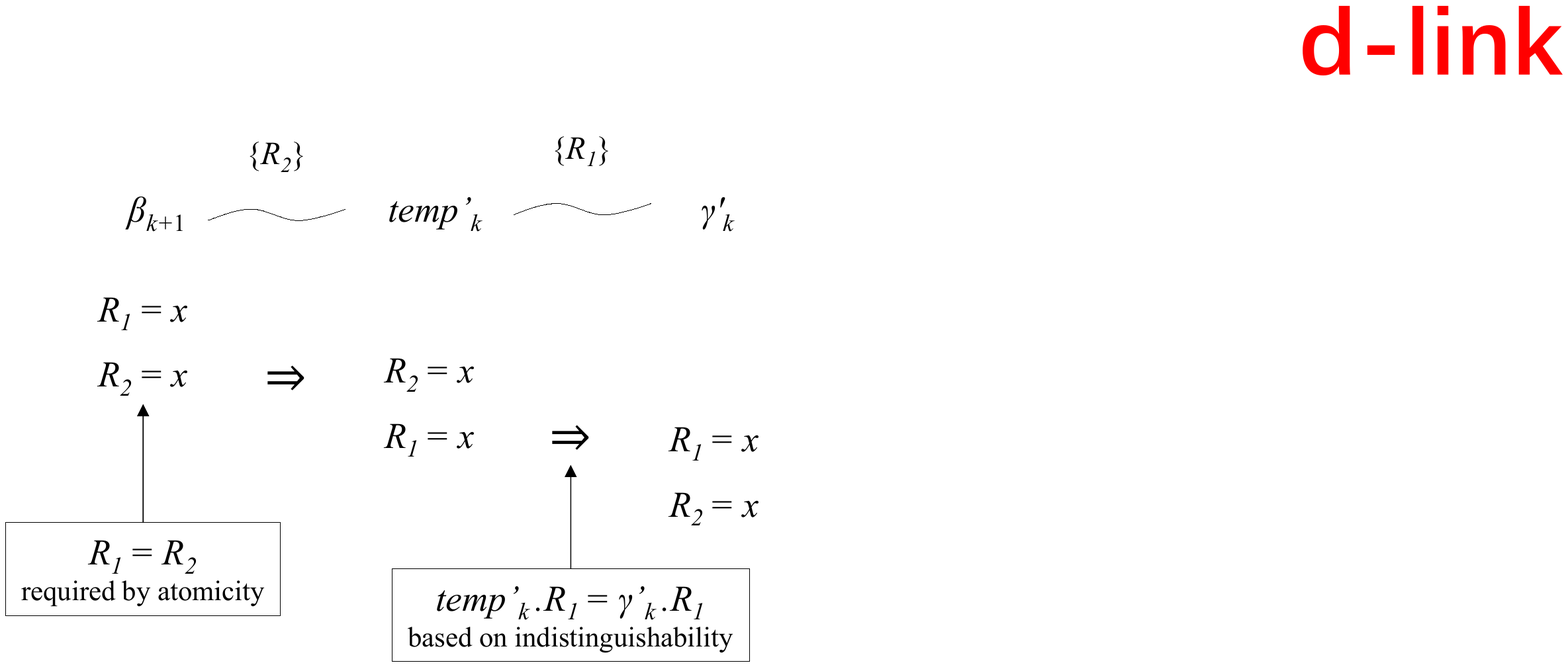}
    \caption{Construction of the diagonal link: the reader's view.}
    \label{F: D-Link1}
\end{figure}

\begin{figure}[ht]
    \center
    \includegraphics[width=\columnwidth]{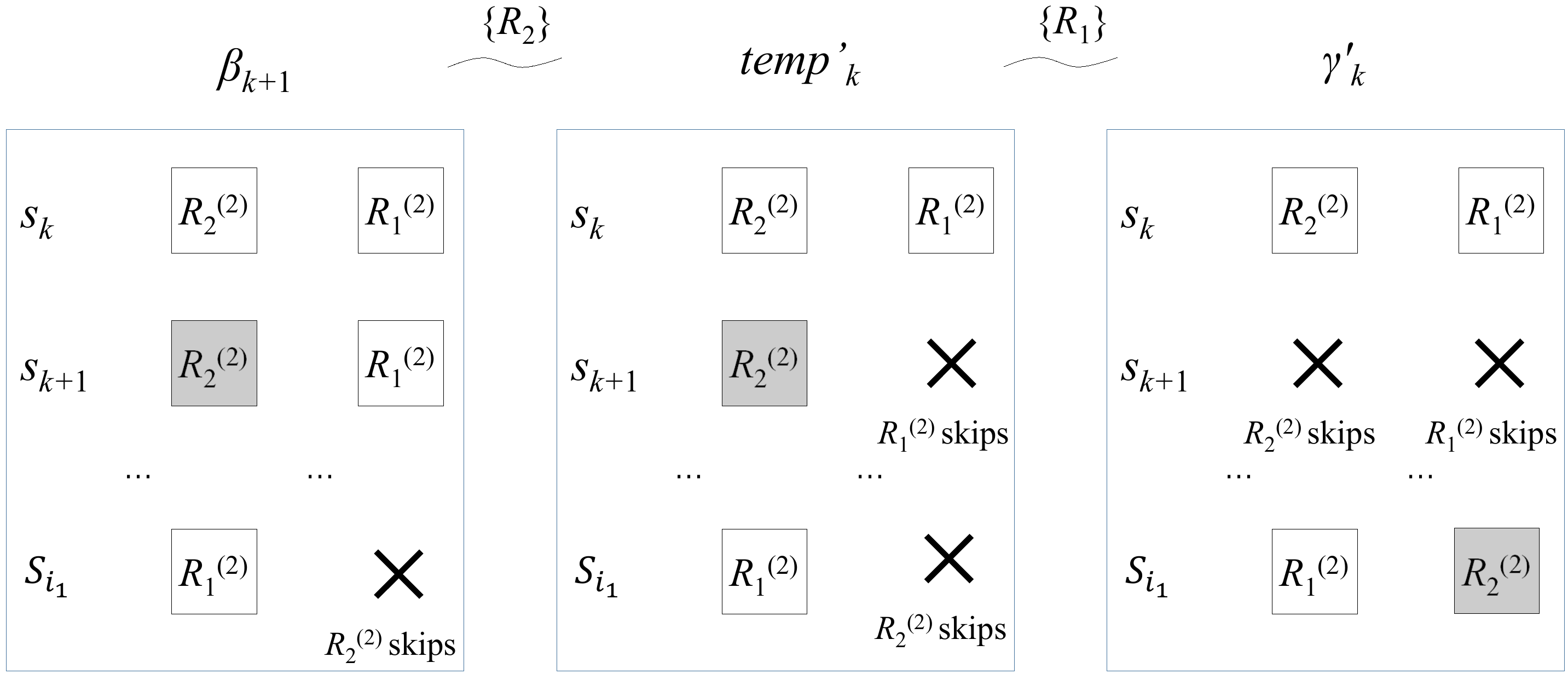}
    \caption{Construction of the diagonal link: the server's view.}
    \label{F: D-Link2}
\end{figure}

First note that in $\beta_{k+1}$, the ``swapping" (see Section \ref{SubSec: Phase2}) takes place in $s_{k+1}$. Thus $s_{k+1}$ sees $R_2^{(2)}$ first and then $R_1^{(2)}$. We create execution $temp_k'$ which is the same with $\beta_{k+1}$, except that $R_1^{(2)}$ skips $s_{k+1}$. The only difference between $temp_k'$ and $\beta_{k+1}$ is on $s_{k+1}$. Since $R_2^{(2)}$ finishes first on $s_{k+1}$, we have that $R_2$ cannot distinguish $\beta_{k+1}$ from $temp_k'$, as shown in Fig. \ref{F: D-Link1}. So $R_2$ will return the same value in $\beta_{k+1}$ and $temp_k'$. As required by the definition of atomicity, $R_1$ will also return the same value in $\beta_{k+1}$ and $temp_k'$. Thus we have $\beta_{k+1} \approx temp_k'$. The construction from the server's view is shown in Fig. \ref{F: D-Link2}.

Now we construct execution $\gamma_k'$, which is the same with $temp_k'$ except that $R_2^{(2)}$ skips $s_{k+1}$ and does not skip $s_{i_1}$ (see Fig. \ref{F: D-Link2}). Similar to the horizontal link case, here we assume that $k+1 \neq i_1$. We will discuss the simpler case ``$k+1=i_1$" below. We need to show that $R_1$ cannot distinguish $temp_k'$ from $\gamma_k'$. The differences concern two servers $s_{k+1}$ and $s_{i_1}$:
\begin{itemize}
    \item As for $s_{k+1}$, since $R_1^{(2)}$ skips $s_{k+1}$, $R_1$ will not see the difference that $R_2^{(2)}$ skips $s_{k+1}$.
    
    \item As for $s_{i_1}$, now we add $R_2^{(2)}$ back on $s_{i_1}$. We can add $R_2^{(2)}$ after $R_1^{(2)}$ on $s_{i_1}$. Thus $R_1$ finishes first on $s_{i_1}$, not being able to distinguish $temp_k'$ from $\gamma_k'$.
\end{itemize}

\noindent Thus we have that $R_1$ returns the same value in $temp_k'$ and $\gamma_k'$. As required by the definition of atomicity, $R_2$ will also return the same value in both executions. This gives us that $temp_k' \approx \gamma_k'$. 

It is straightforward to check that behaviors of $R_1$ and $R_2$ on every server, as well as the order among operations, in $\gamma_i$ and $\gamma_i'$ are the same. Thus we have $\gamma_i' \approx \gamma_i$. Note that here we can see the importance of the seemingly unnecessary change from $temp_k$ to $\gamma_k$ (in Section \ref{SubsubSec: H-Link}): letting $R_1^{(2)}$ skip $s_{k+1}$. The ``unnecessary" skipping of $R_1^{(2)}$ in the horizontal link helps us make $\gamma_k$ and $\gamma_k'$ behave principally in the same way. Finally, this gives us $\beta_{k+1} \approx \gamma_k$.

There is still the case ``$k+1 = i_1$" left, which is  also simpler. We create $\gamma_k'$ as follows. In $\beta_{k+1}$, $s_{k+1}$ only receives $R_1^{(2)}$. Let $R_1^{(2)}$ skip $s_{k+1}$, and we will get $\gamma_k'$. Since $R_2$ skips $s_{k+1}$, $R_2$ cannot distinguish $\beta_{k+1}$ from $\gamma_k'$ and will return the same value. As required by the definition of atomicity, $R_1$ will return the same value in $\beta_{k+1}$ and $\gamma_k'$ too. Thus we still have $\beta_{k+1} \approx \gamma_k'$.

All the horizontal and diagonal links finally connects $\beta_0$ and $\beta_S$, meaning that $R_1$ and $R_2$ return the same value in both executions. However, according to our construction of the chains, $R_1$ and $R_2$ return different values in $\beta_0$ and $\beta_S$. This leads to contradiction, which finishes our impossibility proof.

\section{Fast Write (W1R2): Sieve-based Construction of Executions} \label{Sec: W1R2-Sieve}

Informally speaking, it is reasonable to think that the first round-trip of a read operation should not change the information stored on the servers, thus being not able to affect the return values of other read operations. It is because in the first round-trip, the reader knows nothing about what happens on the servers and other clients. It should not ``blindly" affect the servers.

Following the intuition above, we prove that in our chain argument in Section \ref{Sec: W1R2-Chain}, if $R_2^{(1)}$ affects certain servers, such servers cannot affect our chain argument. Thus we sieve all the servers and only those which actually decide the return values of $R_1$ remain. We restrict our chain argument in Section \ref{Sec: W1R2-Chain} to the remaining servers and can still obtain the contradiction.

Before sieving the servers, we need an abstract model which can capture the essence of the interaction between clients and servers in W1R2 implementations. Only with this abstract model can we discuss what the effect is when we say that the server is affected by the first round-trip of a read. In analogy, the role of this abstract model is like that of the decision-tree model, which is used to derive the lower bound of time complexity for comparison-based sorting algorithms \cite{Cormen09}. We name this model the \textit{crucial-info} model and present it in Section \ref{SubSec: Crucial-Info}. With the crucial-info model, we discuss in Section \ref{SubSec: Sieving} how we can eliminate the servers which have no effect on the return values of read operations. We further explain how our chain argument can be successfully conducted on servers that remain.

\subsection{The Crucial-Info Model} \label{SubSec: Crucial-Info}

We first present the \textit{full-info} model, which is the basis for presenting the crucial-info model. When considering an atomic register implementation, we only care about the number of round-trips to complete a read or write operation. To this end, we use a \textit{full-info} model, where the server is designed as an append-only log. The server just append everything it receives from the writers and readers in its log (never deleting any information). The clients can send arbitrary information to the servers. The clients can also arbitrarily modify the information stored on the servers. The server itself and the clients can always check the log to decide what data the server holds in any moment in the execution.

When the client queries information from the server, the server just replies the client with all the log it currently has. When the client obtains the full-info logs from multiple servers, it derives from the logs what to do next, .e.g. deciding a return value or issuing another round-trip of communication. Since we only care about the number of round-trips required in an implementation, we assume that the communication channel has sufficient bandwidth and the clients and servers have sufficient computing power. Implementations following this model are called full-info implementations.

This full-info model is for the theoretical analysis on the lower bound of the number of round-trips. Obviously, full-info implementations can be optimized to obtain practically efficient implementations.
Since no implementation will use less round-trips than the full-info implementation, we only need to prove that there is no W1R2 full-info implementation of the atomic register.
Based on the full-info model, we can refine certain \textit{crucial information} the servers must maintain. Such crucial information must be stored, modified and disseminated among the clients and the servers, as long as the implementation is a correct atomic register implementation. Specifically, in the executions constructed in our impossibility proof (Section \ref{Sec: W1R2-Chain}), when the writer writes the value ``1" to the servers, the server must store the crucial information ``1". Besides this crucial information, the server can store any auxiliary information it needs, but we are not concerned of such non-crucial information. In analogy, in comparison-based sorting, we only record which elements are compared and what the results are in the decision tree. Other information is not of our concern when deriving the lower bound of the time complexity of comparison-based sorting.

When two writers write ``1" and ``2", no matter what the temporal relation between the two write operations is, the server receives the crucial information in certain sequential order, and we store this crucial information as ``12" or ``21". In order to determine the return value, the reader collects the crucial information ``12" or ``21" from no less than $S-t$ servers. According to the definition of atomicity, the reader needs to infer the temporal relation between the two write operations $W_1$ and $W_2$. Then it can decide the return value. In executions we construct in our proof, the only possible relations between $W_1$ and $W_2$ are:
\begin{itemize}
    \item \textsc{Rel1}: $W_1$ precedes $W_2$.
    \item \textsc{Rel2}: $W_1$ is concurrent with $W_2$.
    \item \textsc{Rel3}: $W_2$ precedes $W_1$.
\end{itemize}

\noindent In the executions in chain $\alpha$, $\beta$, $\gamma$ and $\mathbb{Z}$, there are two essential cases for the reader to decide a return value:
\begin{itemize}
    \item If the reader cannot differentiate \textsc{Rel1} (or \textsc{Rel3}) from \textsc{Rel2}, then it must return 2 (or 1).
     
    \item If two readers both see \textsc{Rel2}, they need to coordinate (through the servers) to make sure that they decide the same return value.
\end{itemize}

\noindent In other cases, the reader can obviously decide what it should return. Note that we only have client-server interaction, i.e., the servers do not communicate with other servers and the clients do not communicate with other clients.

Given the crucial-info model, we can now describe how the first round-trip of a read operation $R_i^{(1)}$ affects another read operation $R_j$. When $R_i^{(1)}$ affects $R_j$, $R_i^{(1)}$ must change the crucial information on some servers, while such modified crucial information is obtained by $R_j$. Note that $R_j$ may be affected (i.e., the indistinguishability is broken) since the crucial information it obtains from the servers changes, but $R_j$ could still decide the same return value even if the crucial information has changed.

In the executions in our proof, the reader only needs to derive the temporal relation between $W_1$ and $W_2$. The only crucial information that can be stored on the server is the temporal order between $W_1$ and $W_2$ the server sees. The possible values of the crucial information on the server are ``12" and ``21". The first round-trip of the reader can only affect the server by changing the crucial information from ``12" to ``21" or vise versa, as long as the implementation correctly guarantees atomicity.

Given the crucial-info model, we can explain how we sieve the serves, as well as how the chain argument can be successfully conducted after the affected servers are eliminated.

\subsection{Eliminating the Affected Servers} \label{SubSec: Sieving}

We conduct the sieving when we append $R_2^{(1)}$ to executions in chain $\alpha = (\alpha_0, \alpha_1, \cdots, \alpha_S)$. From $\alpha_0$, we append the second read operation $R_2$, and discuss the effect of $R_2^{(1)}$ on the return value of $R_1$. Now we have three non-concurrent round-trips $R_1^{(1)}$, $R_2^{(1)}$ and $R_1^{(2)}$, as shown in Phase 2 of Fig. \ref{F: Proof-Overview}. We are concerned of what happens to our chain argument proof (in Section \ref{Sec: W1R2-Chain}) if $R_2^{(1)}$ may affect (the crucial information on) some servers and may potentially affect the return value of $R_1$ (more specifically, $R_1^{(2)}$), thus breaking the indistinguishability we try to construct.

\begin{figure}[ht]
    \centering
    \includegraphics[width=\columnwidth]{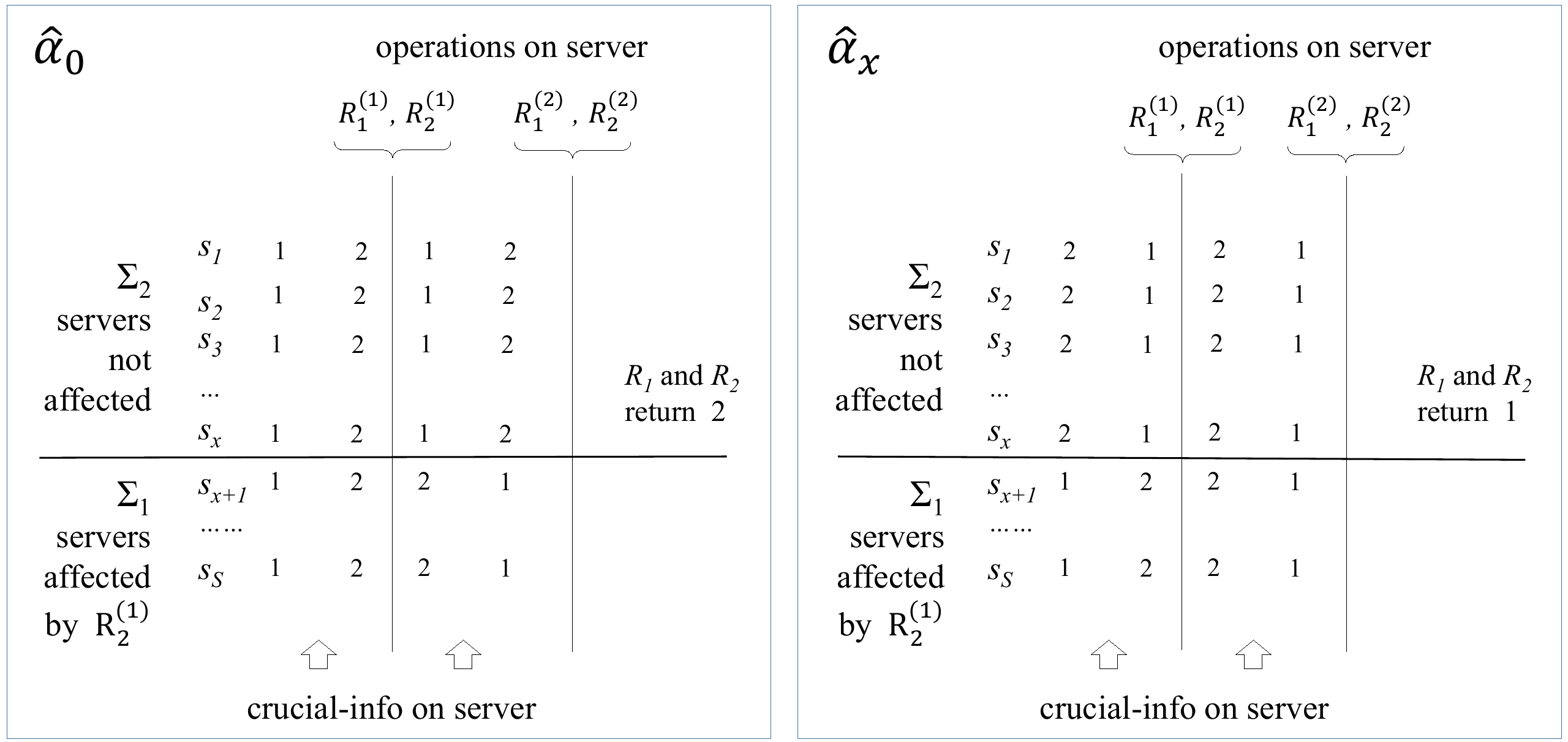}
    \caption{Eliminating servers affected by $R_2^{(1)}$.}
    \label{F: Sieving}
\end{figure}

Considering the effect of $R_2^{(1)}$, we partition all servers $\Sigma_{sv}$ into two subsets $\Sigma_1$ and $\Sigma_2$, as shown in Fig. \ref{F: Sieving}. Set $\Sigma_1$ contains all servers whose crucial information is affected by $R_2^{(1)}$, while $\Sigma_2$ contains all servers whose crucial information is not affected. Without loss of generality, we let $\Sigma_2 = \set{s_1, s_2, \cdots, s_x}$ and $\Sigma_1 = \set{s_{x+1}, s_{x+2}, \cdots, s_S}$. According to our construction of $\alpha_0$, every server in $\Sigma_2$ contains crucial information ``12". At first, the crucial information stored on servers in $\Sigma_1$ is also ``12".  According to the crucial-info model, the only effect on the server which can affect the return value of read operations is changing this ``12" to ``21". So after servers in $\Sigma_1$ are affected by $R_2^{(1)}$, their crucial info is changed from ``12" to ``21". We denote this execution where servers in $S_1$ are affected by $R_2^{(1)}$ as $\hat{\alpha}_0$.

In $\hat{\alpha}_0$, we have that $R_1$ must return 2. It is because $W_1$ precedes $W_2$ by construction, and in any correct atomic register implementation, read operations after $W_2$ should return 2. Whatever the effect of $R_2^{(1)}$ is, it should not prevent $R_1$ from returning 2. For the chain argument, we need to construct the other end of the chain. We still do the swapping one server a time. Execution $\hat{\alpha}_i$ is the same with $\hat{\alpha}_{i-1}$ except for $s_i$, for $1\leq i\leq x$. The crucial information on $s_i$ is ``12" in $\hat{\alpha}_{i-1}$, while the crucial information on $s_i$ is ``21" in $\hat{\alpha}_i$. We do the swapping one server a time for all servers in $\Sigma_2$. The tail execution of chain is $\hat{\alpha}_{tail} = \hat{\alpha}_x$, as shown in Fig. \ref{F: Sieving}. Note that the chain becomes ``shorter". Servers in $\Sigma_1$ are unchanged, in all executions $\hat{\alpha}_0, \hat{\alpha}_1, \cdots, \hat{\alpha}_x$.

Now we describe the sieving process to eliminate servers in $\Sigma_1$ from our chain argument. As for execution $\hat{\alpha}_x$, consider the servers in $\Sigma_1$. They do the computation the same way they do in $\hat{\alpha}_0$, i.e., they first contain crucial info ``12", then is affected by $R_2^{(1)}$ and change their crucial info to ``21". Note that the effect of $R_2^{(1)}$ is ``blind" effect because it does not obtain any information from the outside world first. The servers in $\Sigma_1$ and the reader $r_2$ of round-trip $R_2^{(1)}$ will not differentiate $\hat{\alpha}_x$ from $\hat{\alpha}_0$. Thus all servers in $\Sigma_1$ behave the same way in both executions, and they will have crucial info ``21". 

As for servers in $\Sigma_2$ in $\hat{\alpha}_x$, after $W_1$ and $W_2$, all servers in $\Sigma_2$ have crucial information ``21". This crucial information should remain ``21" after $R_1^{(1)}$ and $R_2^{(1)}$. Assume for contradiction that the crucial information on some server $s_y$ in $\Sigma_2$ has been affected by $R_1^{(1)}$ and $R_2^{(1)}$, and is changed from ``21" to ``12". Combining the behavior of $s_y$ in both $\hat{\alpha}_0$ and $\hat{\alpha}_x$, we find that $s_y$ always end with crucial information ``12" after $R_1^{(1)}$ and $R_2^{(1)}$, no matter what the write operations write on the servers. Such servers obviously cannot decide the return value of $R_1$ and can be safely eliminated. So we can assume that all servers in $\Sigma_2$ in $\hat{\alpha}_x$ have crucial information ``21" after $R_1^{(1)}$ and $R_2^{(1)}$.

In this way, $R_1$ will see all servers have crucial info ``21" in $\hat{\alpha}_x$, and $R_1$ must return 1 in execution $\hat{\alpha}_x$. We thus obtain the key property required for the chain argument: in two end executions of the chain $\hat{\alpha} = (\hat{\alpha}_0, \hat{\alpha}_1, \cdots \hat{\alpha}_x)$, $R_1$ return different values. Note that the length of the chain will not affect our chain argument in Section \ref{Sec: W1R2-Chain}, as long as we have enough servers left for the chain argument. Operation $R_1$ uses crucial information only from servers in $\Sigma_2$. Crucial information on servers in $\Sigma_1$ have been affected, and the change in this crucial information will not affect that $R_1$ returns 2. Since $t=1$ and servers in $\Sigma_2$ can enable a correct atomic register implementation (we have this assumption to derive the contradiction), we have at least 3 servers in $\Sigma_2$.

Another threat to clarify is that when constructing chain $\beta'$, $\beta''$ and $\beta$, the chains are based on the swapping among all servers, i.e., chain $\beta'$, $\beta''$ and $\beta$ all have length $S+1$ even after the sieving. That is to say, the sieving is only conducted on executions in chain $\alpha$, in order to obtain the critical server $s_{i_1}$. This raises the potential threat that when constructing $\beta_0, \beta_1, \cdots, \beta_S$, what happens if $R_1^{(1)}$ affects the return value of $R_2$. Observe that in our proof, we only use the fact that, when $R_2$ (both round-trips) skips $s_{i_1}$ in executions $\beta'_S$ and $\beta''_S$, $R_2$ returns the same value. This means that, no matter what the effect of $R_1^{(1)}$ is, $R_2$ still returns the same values in both $\beta'_S$ and $\beta''_S$, as long as the critical server is skipped. Thus our chain argument can successfully go on as in Section \ref{Sec: W1R2-Chain}.

\section{Fast Read (W2R1): Impossibility and Implementation} \label{Sec: W2R1}

In this section, we discuss the impossibility and implementation of fast read (W2R1) multi-writer atomic registers. The necessary and sufficient condition of a W2R1 implementation is $R< \frac{S}{t}-2$, which is the same with that of the single-writer case \cite{Dutta10}. The impossibility proof and the algorithm design are also obtained by extending their counterparts in the single-writer case.

\subsection{Impossibility when $R \geq \frac{S}{t}-2$}

We need to prove that it is impossible to obtain a W2R1 implementation when $R \geq \frac{S}{t}-2$ in the multi-writer case. It is sufficient to prove that, even there is only one writer and this single writer can employ two round-trips, W2R1 implementations are still impossible.

\begin{figure}[ht]
    \centering
    \includegraphics[width=0.5\columnwidth]{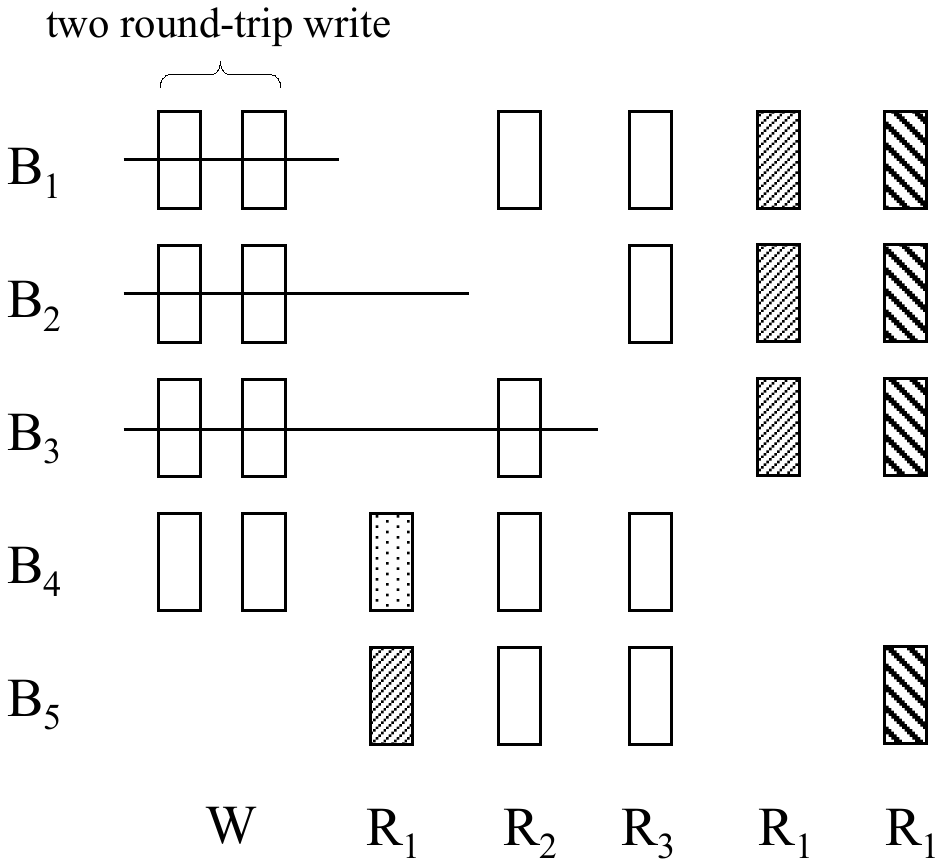}
    \caption{Fast read impossibility.}
    \label{F: Dutta-Sieve}
\end{figure}

The proof in the single writer-case does not depend on how many round-trips a write operation has. When we change all write operations in the impossibility proof in the single-writer case to two (or more) round-trips, we let all the two (or more) round-trips of a write operation take place consecutively and precede all other operations, as shown in Fig. \ref{F: Dutta-Sieve} (based on Fig. 6 of \cite{Dutta10}). The rest of the impossibility proof is not affected. 

\subsection{Implementation when $R < \frac{S}{t}-2$}

We derive the W2R1 implementation from the single-writer W1R1 implementation in \cite{Dutta10}. Our implementation is inspired by how multiple writers are handled in the W2R2 implementation \cite{Lynch97}, which can also be viewed as a derivation from the single-writer W1R2 implementation \cite{Attiya95}. The key change in the design of a multi-writer implementation is that we use $(ts, w_i)$ to denote one value. Here $w_i$ is the writer ID and $ts$ is the version number denoting one value written by $w_i$. Assuming that the writer IDs are totally ordered, we can thus order all the values from multiple writers using the lexicographical order when we have equal $ts$ values.

The order among write values is further strengthened by the two round-trip write algorithm. Specifically, before writing a value, the writer first queries all the servers and calculates the $maxTS$ in its first round-trip. Then the writer updates value $(maxTS+1, w_i)$ to all servers in the second round-trip. The two-round-trip write algorithm guarantees that when write operations have the same $ts$ value, they must be concurrent.

As for the the single round-trip read operations, the reader first obtains multiple values from the servers. It also uses the $admissible(\cdot)$ predicate (defined in the single-writer algorithm in \cite{Dutta10}) to test all the values obtained. The $admissible(\cdot)$ predicate is designed to guarantee that: i) a read never returns older values than that of a preceding write, and that ii) a read never returns older values than that of a preceding read. Since there are multiple writers, the reader may obtain multiple admissible values and need to return one of them. Since all the values are totally ordered, we simply let the reader return the largest admissible value. That is, when the reader needs to choose from equal $ts$ values, it just chooses the $ts$ value with the largest writer ID.

One potential threat to the correctness of our algorithm is that in the single-writer case, all values are totally ordered on one single-writer and it is trivial to choose the more up-to-date value. However in the multi-writer case, the two round-trip algorithm can order non-concurrent write operations from different writers. But for the concurrent writes, we can only use the (somewhat arbitrary) order among writer IDs. We need to prove that using the writer ID order will not comprise the correctness of our implementation.

Specifically, for two read operations $R_1$ preceding $R_2$,  the predicate $admissible(\cdot)$ guarantees that $S_{ad}(R_1) \subseteq S_{ad}(R_2)$ \cite{Dutta10}. Here $S_{ad}(R_i)$ ($i=1, 2$) denotes the set of admissible values on $R_i$. Denote the return values of $R_1$ and $R_2$ as $val_1 = \max(S_{ad}(R_1))$ and $val_2 = \max(S_{ad}(R_1))$ respectively. The potential threat to our multi-writer implementation is that $R_2$ chooses a new return value only due to the difference in writer ID while the $ts$ values are the same. Specifically, assume that $val_2 \neq val_1$, but $val_2.ts = val_1.ts$ and $val_2.writer\text{-}id > val_1.writer\text{-}id$. Since $val_2 \not \in S_{ad}(r_1)$ (or $R_1$ and $R_2$ will choose the same return value), we have that $val_2$ is not admissible in $R_1$'s view, but $val_2$ is admissible in $R_2$'s view. This ensures that $R_1$ must be concurrent with $W_2$ (let $W_i$ denote the write operation of $val_i$ for $i=1, 2$). Thus we have $W_1, R_1, W_2, R_2$ is a correct permutation of these operations as required by the definition of atomicity. This ensures that the return value of $R_2$ is correct.

For other cases, the correctness proof of our W2R1 implementation is principally the same with that of the W1R1 implementation in \cite{Dutta10}. We present our W2R1 implementation and its detailed proof of correctness in Appendix \ref{Appen: Proof}.
Note that impossibility results in the crash failure model directly imply impossibility in the Byzantine failure model. However, for our W2R1 implementation, we can further study whether it can be extended to further tolerate Byzantine failures. The extension is principally the same with that in the single-writer case, as detailed in \cite{Dutta10}. We thus omit detailed discussions here.

\section{Related Work} \label{Sec: RW}


The importance of low latency data access in distributed storage systems motivates the study on fast implementations of distributed atomic registers. Fast implementation in the single-writer case is studied in \cite{Dutta10}, where the sufficient and necessary condition for fast implementation is derived. As for the multi-writer case, only impossibility for fast read-write implementations is presented. When examined at a finer granularity, it is still open whether fast implementations are possible when only read or write are required to be fast. The notion of semifast implementation is presented in \cite{Georgiou09}. It is proved that semifast implementation is not possible for multi-writer atomic registers. In this work, we consider implementations where the read can always be slow (using two or more round-trips). The implementation we consider is strictly stronger than semifast implementations. Thus our our impossibility proof is more general and directly implies the impossibility of semifast implementations. This work concludes this series of studies on fast implementations of distributed atomic registers. Impossibility proof for fast write implementations is presented, and necessary and sufficient condition for fast read implementations is derived.

Our impossibility proof for fast write implementations are inspired by the classical result in  a shared-memory setting that ``atomic reads must write" \cite{Lamport86a, Lamport86b, Attiya04}. The CAP theorem \cite{Brewer00, Gilbert12} and the PARCELC tradeoff \cite{Abadi12} in distributed systems also inspire us to prove the impossibility of fast (low latency, strongly consistent and fault-tolerant) implementations. Our use of the crucial-info model is inspired by the CHT proof of the weakest failure detector for consensus \cite{Chandra96weakest, Freiling11}. In the CHT proof, a directed acyclic graph is used to store the  failure detector outputs on all processes as well as the temporal relations between them.

The study on atomic register implementations on the Oh-RAM model is closely related to our work \cite{Hadjistasi17}. Both works use chain arguments \cite{Attiya14} to prove the impossibility. The main difference lies in the system model. In the Oh-RAM model, servers are allowed to exchange messages , while in our client-server model, we only model communications between the client and the server. We derive our system model from the existing work \cite{Attiya95, Lynch97, Aspnes19, Dutta10}, as well as from our study of popular distributed storage systems \cite{Lakshman10, Redis20, Riak20}.

\section{Conclusion and Future Work} \label{Sec: Concl}

In this work, we study fast write and fast read implementations of multi-writer atomic registers. For fast write implementations, we come up with the impossibility proof, which is based on a three-phase chain argument. For fast read implementations, we provide the necessary and sufficient condition for fast implementations, by extending the result of the single-writer case.

In our future work, we will study fast implementations for multi-writer atomic registers from a different perspective. Specifically, we will fix fast implementations in the first place, and then quantify how much data inconsistency will be introduced when strictly guaranteeing atomicity is impossible. We also plan to introduce knowledge calculus to reason about quorum-based distributed algorithms at a higher level of abstraction. 

\bibliographystyle{acm}
\bibliography{arxiv}

\clearpage

\appendix

\section{Correctness Proof of the W2R1 Implementation} \label{Appen: Proof}

\subsection{Definitions}

An execution satisfies atomicity if for every history $H'$ of any of it there is a history $H$ that completes $H'$ and $H$ satisfies the properties below. Let $\Pi$ be the set of all operations in $H$. There is an irreflexive partial ordering $\prec_{\pi}$ of all the operations in $H$ such that (A1) if $op_1$ precedes $op_2$ in $H$, then
it is not the case that $op_2 \prec_{\pi} op_1$; (A2) if $op_1$ is a write operation in $\Pi$ and $op_2$ is
any other operation(including other write operation) in $\Pi$, then either $op_2 \prec_{\pi} op_1$ or $op_1 \prec_{\pi} op_2$ in $\Pi$; and (A3) the value returned by each read operation is the value written by the last preceding write operation according to $\prec_{\pi}$.

In a given execution, we denote by $wr_{k,i}$ the write that is preceded by exactly the write with $ts = k$ by the writer $w_i$ in the execution (note that $wr_{0,\perp}$ is not invoked by the writer). Then we say that an operation $op$ returns a value $(k,w_i)$, if (a) $op$ is $wr_{k,i}$, or (b) $op$ is a read that returns the value stored by $wr_{k,i}$.

For two values $(ts_1, w_i)$ and $(ts_2, w_j)$, we say $(ts_1, w_i) < (ts_2, w_j)$ if and only if $(ts_1 < ts_2) \vee (ts_1 = ts_2 \wedge w_i < w_j)$. Consider a relation $\prec_{\pi}$ such that $op_1 \prec_{\pi} op_2$ if and only if the value returned by $op_1$ is smaller than the value returned by $op_2$. Then it is straightforward to show that the $\prec_{\pi}$ is a partial ordering that satisfies properties A1 - A3 if the following properties are satisfied:\\
(MWA0) Let $wr$ and $wr'$ be two different write operations, and $v_1$(resp., $v_2$) is the value that $wr$ (resp., $wr'$) writes (note that $v_1 \neq v_2$). If $wr \prec_{\sigma} wr'$, then $v_1$ $<$ $v_2$.\\
(MWA1) If a read returns, it returns a nonnegative timestamp and a $wid$ of the writer proposing the timestamp.\\
(MWA2) If a read $rd$ returns value $(l,w_j)$ and $rd$ follows write $wr_{k,i}$, then $(l, w_j) \geq (k,w_i)$.\\
(MWA3) If a read $rd$ returns value $(k, w_i)$, then $rd$ does not precede $wr_{k,i}$.\\
(MWA4) If reads $rd_1$ and $rd_2$ return value $(k,w_i)$ and $(l,w_j)$, respectively, and if $rd_2$ follows $rd_1$, then $(l,w_j) \geq (k,w_i)$.\\

\subsection{Implementation and Correctness of Algorithm 1 and 2}

In \cite{Dutta10}, there is a W1R1 atomic register implementation for the single-writer case. We use this implementation for reference and extend it to a W2R1 implementation, as shown in Algorithm 1 and 2. Since Algorithm 1 and 2 largely reuse the implementation in \cite{Dutta10}, we only give the pseudo code and do not explain them in detail. Now we prove the correctness of Algorithm 1 and 2. Our proof also largely reuses the proof in \cite{Dutta10}.

\begin{algorithm}[ht] 
    \label{Alg-client}
    
    \IncMargin{1em}
    \DontPrintSemicolon
    \caption{Client logic}
    \w(){} \;
    \proc(initialization:){
        $ts$ $\leftarrow$ $0$ \;
    }	
    \proc(\writers{}){
        send($read$, $maxTS$) to all servers \;
        Wait until receive $\text{READACK}$ from $S-t$ servers  \;
        $maxTS$ $\leftarrow$ Max $\{$$ts$ $in$ $\text{READACK}s$ $\}$ \;
        $ts$ $\leftarrow$ $maxTS + 1$ \;
        $val$ $\leftarrow$ ($ts$, $w_i$) \;
        send($write$, $val$) to all servers \;
        Wait until receive $\text{WRITEACK}$  from $S - t$ servers \;
        \Return{$\emph{OK}$} \;
    }
    \BlankLine
    \BlankLine
    \BlankLine
    
    \r(){} \;
    \proc(initialization:){
        $valQueue$ $\leftarrow$ $(0,\perp)$ \;
    }	
    \proc(\readers{}){
        send($read$, $valQueue$) to all servers \;
        Wait until receive $\text{READACK}$ from $S-t$ servers  \;
        $rcvMsg$ $\leftarrow$ $\{$ $m$ | $r_i$ received $\text{READACK}$ $\}$ \;
        $valQueue$ $\leftarrow$ $(\bigcup_{v \in rcvMsg} v)$ $\cup$ $valQueue $ \;
        $maxV$ $\leftarrow$ Max $\{$ ($ts$,$wid$) in $rcvMsg$ $\}$ \;
        \while(){
            \If {$\exists$ a $\in$ $[1, R+1]$ : $admissible(maxV, rcvMsg, a)$}
            {
                \Return $maxV$ \;
            }
            \Else
            {
                remove $maxV$ from all msg in $rcvMsg$\;
                $maxV$ $\leftarrow$ Max $\{$ ($ts$,$wid$) in $rcvMsg$ $\}$ \;
            }
        }
    }
    \BlankLine
    $admissible(v,Msg,a)$ $\equiv$ $\exists \mu \subseteq Msg \forall m \in \mu:$ ($m$ has $v$) $\wedge$ ($|\mu|$ $\geq$ $S-at$) $\wedge$ ($|\bigcap_{m'\in \mu}m'.updated| \geq a$)
    
    \BlankLine
    \BlankLine
    \BlankLine
\end{algorithm}

\begin{algorithm}[ht] 
    \label{Alg-server}
    \IncMargin{1em}
    \DontPrintSemicolon
    \caption{Server logic}
    
    \s(){} \;
    \proc(initialization:){
        $val_i$ $\leftarrow$ $(0,\perp)$ \;
        $value_{vector}$ $\leftarrow$ $\{val_i, update\}$ \;
        $value_{vector}[val_i].update$ $\leftarrow$ $\emptyset$ \;
    }
    \proc(\update{$val,c$}){
        \If{$val$ > $val_i$}
        {
            $value_{vector}$ $\leftarrow$ $value_{vector}$ $\cup$ $\{val, update\}$ \;
            $value_{vector}[val].updated$ $\leftarrow$ $\{$ $c$ $\}$ \;
            $val_i$ $\leftarrow$ $val$ \;
        }
        \Else{
            $value_{vector}[val].updated$ $\leftarrow$ $value_{vector}[val].updated$ $\cup$ $\{$ $c$ $\}$ \;
        }
    }
    \upon({\receive $(write, val)$ from writer $w_k$})
    {
        \update{$val$, $w_k$} \;
        send $\text{WRITEACK}$ to $w_k$ \;
    }
    \upon({\receive $(read, valQueue)$ from reader $r_j$})
    {
        \update{$val$, $r_j$} for all $val$ in $valQueue$ \;
        send $\text{READACK}$ to $r_j$ \;
    }
\end{algorithm}

To prove that Algorihtm 1 and 2 implement an atomic register, it suffices to prove MWA0-MWA4. In the proof, we use the following notations.

\paragraph{Definition 1} $rcvMsg_{op}$ denotes the set of received $\text{READACK}$ messages the reader collects in read operation $op$.

\paragraph{Definition 2} $\Sigma_{op}$ denotes the set of servers from which the reader received $\text{READACK}$ messages in $rcvMsg_{op}$ (in case $op$ is a read), or the set of servers from which the writer received $\text{WRITEACK}$ messages of $op$ (in case $op$ is a write). Notice that for every operation $op$, $|\Sigma_{op}| = S-t$.

\paragraph{Definition 3} $maxV^{old}_{op}$ denotes the value a read $op$ send to all servers when it starts reading. Moreover, $maxV_{op}$ denotes the value computed by the reader in line 23 in Algorithm 1, in $op$. $maxTS^{old}_{op}$ denotes the timestamp in $maxValue^{old}_{op}$ and $maxTS_{op}$ denotes the timestamp in $maxValue_{op}$.

\paragraph{Definition 4} $\mu_{op,v,a}$ denotes, in case value $v$ is admissible with degree $a$ in $op$, the subset of $rcvMsg_{op}$, such that

(a) $|\mu_{op,v,a}| \geq S - at$,

(b) for all $m \in \mu_{op,v,a}$, $m$ has $v$, and

(c) $|\bigcap_{m \in \mu_{op,v,a}} m.updated| \geq a$.

\paragraph{Definition 5} $\Pi_{op,v,a}$ denotes the set of servers $\bigcap_{m \in \mu_{op,v,a}} m.updated$.

\paragraph{Definition 6} $\Sigma_{op,v,a}$ denotes the set of servers that sent messages in $\mu_{op,v,a}$.

We start with several simple observations that we use in the rest of the proof.

\paragraph{Lemma 0 (MWA0)} Let $wr$ and $wr'$ be two different write operations, and $v_1$(resp., $v_2$) is the value that $wr$ (resp., $wr'$) writes (note that $v_1 \neq v_2$). If $wr \prec_{\sigma} wr'$, then $v_1$ $<$ $v_2$.

Proof. The lemma trivially holds by the definition of value. \pfend

\paragraph{Lemma 1} If a server gets variable $x$ at time $T$ , then the server send all $\text{ACK}$ with $x$ after time $T$ .

Proof. The lemma is proved by trivial server code inspection. \pfend

\paragraph{Lemma 2} Read operation $rd$ may only return a value whose timestamp is either $maxTS_{rd}$ or $maxTS_{rd}-1$.

Proof. $rd$ will get $maxV_{rd} = k$ in several servers. Since $wr_{k,i}$ starting proposing $maxV_{rd}$, $w_i$ must have finished proposing ($k - 1$, $w_i$). So ($k - 1$, $w_i$) will be admissible, and $rd$ will return a value larger than $(k-1,w_i)$. \pfend

\paragraph{Lemma 3} $maxV^{old}_{rd} $ is admissible in $rd$.

Proof. Recall that $maxV^{old}_{rd}$ denotes the value sent in read message in $rd$ (line 19 in Algorithm 1). By lines 21 in Algorithm 1, every $\text{READACK}$ message received by a reader in $rd$ from some server $s_j$ has value $maxV^{old}_{rd}$. Hence, $maxTS_{rd} \geq maxTS^{old}_{rd}$ . Since $\Sigma_{op} = S - t$, so $maxV^{old}_{rd} $ is admissible with degree $a = 1$ in $rd$. \pfend

Using these several simple observations, we can first prove MWR1 and MWR2.

\paragraph{Lemma 4 (MWA1)} If a read returns, it returns a value with nonnegative timestamp.

Proof. To prove the lemma, it is sufficient to show that there is no read $rd$ in which $maxTS^{old}_{rd} < 0$ (then the lemma follows from Lemma 3).

To see this, assume by contradiction that there is a read $rd$ by reader $r_i$ in which $maxTS^{old}_{rd} < 0$. By lines 17 in Algorithm 1, this is not the first read by $r_i$, i.e., there is a read $rd'$ by $r_i$ that (immediately) precedes $r_i$ such that $maxTS_{rd} < 0$, i.e., $S-t$ servers sent $\text{READACK}$ messages in $rd'$ with $ts' < 0$. However, since server timestamps are initialized to 0, this contradicts Lemma 2. \pfend

\paragraph{Lemma 5 (MWA2)} let a read $rd$ which returns value $(l,w_j)$ follow write $wr_{k,i}$, then $(l, w_j) \geq (k,w_i)$.

Proof: Denote by $r_i$ the reader that invoked $rd$ and let $\Sigma' = \Sigma_{wr_{k,i}} \cap \Sigma_{rd}$. Since $|\Sigma_{wr_{k,i}}| = S-t$ and $|\Sigma_{rd}| = S-t$, we have $|\Sigma'| \geq S - 2t$.

When a server $s_j$ in $\Sigma_{wr_{k,i}}$ (and, hence, in $\Sigma'$) replies to a write message from $wr_{k,i}$ in time $T$, $(k,w_i)$ will be in all messages $s_j$ sent to clients after $T$. Since $wr_{k,i}$ precedes $rd$, $rd$ will see $(k,w_i)$ in all servers in $\Sigma'$. So $(k,w_i)$ will be admissible with degree $a = 2$ in $rd$. And so $rd$ must return a value $(l,w_j)$ larger than $(k,w_i)$. \pfend

Then we need to prove MWA3. The following lemma helps prove property MWA3.

\paragraph{Lemma 6} If $maxV_{rd} \geq (k,w_i)$, then $rd$ does not precede $wr_{k,i}$.

Proof. We focus on the case $k \geq 1$, since the proof for $k = 0$ follows from the definition of $wr_{o,\perp}$. To prove the lemma, it is sufficient to show that no server has value $(k,w_i)$ before $wr_k$ is invoked. Assume, by contradiction, that there is such a server $s_i$ that is, moreover, the first server to has a value greater than $(k,w_i)$ according to the global clock (at time $T$ ); i.e., no server has a value $(l,w_j)$ greater than $(k,w_i)$ before time $T$. It is obvious that $wr_{k,i}$ is invoked atfer $T$; and so is the second round trip of $wr_{l,j}$ (which might be the same operations), or $(k,w_i)$ will greater than $(l,w_j)$. Hence, $s_i$ must have get $(k,w_i)$ after receiving a read message in a read $rd'$ invoked by reader $r_x$ in which has value $(k,w_i)$. Since $(l, w_j) \geq (k, w_i) > (0,\perp)$, there is a read $rd''$ by $r_x$ that immediately precedes $rd'$ which has receive a message containing $(k,w_i)$. Since $rd''$ precedes $rd'$, $rd''$ completes before time $T$ . So some server had got value $(l,w_j)$ before $rd''$ completed. A contradiction with the assumption that no server gets value $(l,w_j)$ before time $T$. \pfend

Lemma 6 has the following important corollary.

\paragraph{Corollary 1} If $maxV_{rd} = (l,w_j) > (0,\perp)$, all $wr_{k,i}$($k < l$) completes before $rd$ completes.

\paragraph{Lemma 7 (MWA3)} If a read $rd$ returns value $(k, w_i)$, then $rd$ does not precede $wr_{k,i}$.

Proof. The lemma is proved by Lemmas 2 and 6.
By lemma2, $rd$ will only return value with timestamp $maxTS_{rd}$ or $maxTS_{rd} - 1$. If $k = maxTS_{rd} - 1$, then $(k,w_i) < maxV_{rd}$; by corollary 1, $wr_{k, x}$ proceeds $rd$. If $k = maxTS_{rd}$, by lemma 6, $rd$ does not proceed $wr_{maxV_{op}}$; and it is obvious that $wr_{maxV_{op}}$ (which may be the same $wr$ as $wr_{maxV_{op}}$) does not proceed $(maxTS_{rd},k)$; so $rd$ does not proceed $wr_ {k,i}$. \pfend

\paragraph{Lemma 8} $v$ is admissible in $rd_1$ and $rd_2$ follows $rd_1$, then $v$ is admissible in $rd_2$.

Proof. $ret_{rd_1}$ must be admissible in $rd_1$ with degree a. There are two cases:

(i) $a \leq R$.
$v$ is admissible in $rd_1$ and is not admissible in $rd_2$. We show that this case is impossible by exhibiting appropriate contradictions. In this case, by Lemma 10, there is at least one server $s_i \in \Sigma_{\mu_{rd_1,v,a}} \bigcap \Sigma_{rd_2}$. Since $rd_1$ precedes $rd_2$, $s_i$ first replies with $v$ to $rd_1$ before $s_i$ replies to $rd_2$. Finally, by Lemma 1, it follows that $s_i$ replies to $rd_2$ with $v$. Let $\mu_1$ be the set of $\text{READACK}$ messages sent from servers in $\Sigma_{rd_1,v,a} \bigcap \Sigma_{rd_2}$ to $rd_1$. Denote $\bigcap_{m\in \mu_1} m.v.updated$ by $\Pi_1$. Notice that, by definitions of $\mu_1$ and $\mu_{rd_1,v,a}$, $\mu_1 \subseteq \mu_{rd_1,v,a}$. Hence, we have $\Pi_{rd_1,v,a} \subseteq \Pi_1$ and $|\Pi_1| \geq a$. Let $\mu_2$ be the set of messages received by $rd_2$ from servers in $\Sigma_{\mu_{rd_1,v,a}} \bigcap \Sigma_{rd_2}$. For any server $s_i \in \Sigma_{\mu_{rd_1,v,a}} \bigcap \Sigma_{rd_2}$, let $m_1$ and $m_2$ be the messages sent by $s_i$ in $\mu_1$ and $\mu_2$, respectively. Since $m_1$ is sent before $m_2$, we have $m_1.v.updated \subseteq m_2.v.updated.$ Hence, $\Pi_1 \subseteq \bigcap_{m\in \mu_2} m.v.updated$. Since every server which replies to $r_2$ in $rd_2$ adds $r_2$ to its updated set before replying to $r_2$, $r_2 \in \bigcap_{m \in \mu_2} m.v.updated$. Since $r_2 \notin \Pi_1, |\bigcap_{m \in \mu_2} m.v.updated| \geq |\Pi_1| + 1 \geq a + 1$. Since (a) the number of messages in $\mu_2$ equals the number of servers in $\Sigma_{rd_1,a} \bigcap \Sigma_{rd_2}$, and (b) $a + 1 \leq R + 1$, by Lemma 10 and the definition of predicate admissible, we have that $v$ is admissible in $rd_2$ with degree $a+1$.

(ii) $a = R + 1$.
Since $|{w, r_1, \cdots , r_R}| = R + 1$ and $|\bigcap_{m \in \mu_{rd_1,v,a}} m.updated| \geq a = R+1$, we have $r_2 \in \bigcap_{m \in \mu_{rd_1,v,a}} m.updated$. By Lemma 9, $\Sigma_{\mu_{rd_1,v,a}}$ contains at least $t+1$ servers. Let $rd_2'$ be the last read by reader $r_2$ which precedes $rd_1$.  Since $|\Sigma_{rd}| = S-t$, there is at least one server $s_k$ in $\Sigma_{\mu_{rd_1,v,a}} \cap \Sigma_{rd_2'}$, such that the $\text{READACK}$ message $m$ sent by $s_k$ is received by $r_2$ in $rd_2'$. In the following paragraph, we show that $m$ contains $ret_{rd_1}$.

By contradiction, assume $m$ does not contain $ret_{rd_1}$. There is a read $rd_{\alpha}$ by $r_2$ , such that $rd_{\alpha}$ follows $rd_2'$ and $s_k$ sends a $\text{READACK}$ message $m_{\alpha}$ to $rd_{\alpha}$, before $s_k$ sends $m_k \in \mu_{rd_1,v,a}$, i.e., before $rd_1$ is invoked. Hence, $rd_2'$ is not the last read by reader $r_2$ which precedes $rd_1$ - a contradiction.

Since $m$ contains $ret_{rd_1}$ and $rd_2$ follows $rd_2'$, it follows that $r_2$ in $rd_2$ sends read messages with $value \geq maxValue_{rd_1}$. So all servers in $\Sigma_{rd_2}$ will send $v$ to $rd_2$, so $v$ is admissible in $rd_2$ with degree $a = 1$. \pfend

Lemma 8 has the following important corollary, which is MWA4.

\paragraph{Corollary  2  (MWA4)} If reads $rd_1$ and $rd_2$ return value $(k,w_i)$ and $(l,w_j)$, respectively, and if $rd_2$ follows $rd_1$, then $(l,w_j) \geq (k,w_i)$. \\

The following 2 auxiliary lemmas are related to the predicate admissible and to the sizes of the relevant subsets of the set $rcvMsg$.

\paragraph{Lemma 9} If $maxV_{rd}$ $v$ is admissible in rd with degree a, then $\Sigma_{\mu_{rd,v,a}}$ contains at least $t + 1$ servers.

Proof. By Definition 1 - 6 and inequalities $a \leq R + 1$ and $R < \frac{S}{t} - 2$, we have 

\begin{center}
$|\Sigma_{\mu_{rd,v,a}}| \geq S - at > (R+2)t - (R+1)t = t$.  \pfend
\end{center}

\paragraph{Lemma 10} Assume that $maxV_{rd}$ is admissible with degree $a \in [1,R + 1]$ in some read $rd$ and that a complete read $rd'$ follows $rd$. Then the number of servers in $\Sigma_{\mu_{rd,v,a}} \cap \Sigma_{rd}$ is at least $S - (a + 1)t$. Moreover, $\Sigma_{\mu_{rd,v,a}} \cap \Sigma_{rd}$ contains at least one server.

Proof. Since $|\Sigma_{\mu_{rd,v,a}}|=|\mu_{rd,v,a}| \geq S-at$ and $|\Sigma_{rd}| = S-t$, it follows that $\Sigma_{\mu_{rd,v,a}} \cap \Sigma_{rd} \geq S - (a+1)t$. Moreover, since $a \in [1,R + 1]$ and $t < \frac{S}{R+2}$, we have $S - (a + 1)t \geq 1$. \pfend

\end{document}